\documentclass[12pt]{article}
\usepackage{epsfig}
\usepackage{cite}
\usepackage{color}
\usepackage{comment}
\usepackage{color}

\def\bea{\begin{eqnarray}}
\def\eea{\end{eqnarray}}
\def\bq{\begin{quote}}
\def\eq{\end{quote}}
\def\nn{\nonumber}
\newcommand{\bqa}{\begin{eqnarray}}
\newcommand{\eqa}{\end{eqnarray}}
\newcommand{\psb}{\bar{\psi}}

\begin{document}
\pagestyle{empty}

\begin{flushleft}
{\bf December 12, 2005}\\
{\tt IFJPAN-V-2005-11}
\end{flushleft}
\vspace*{2cm}
\begin{center}
{\Large \bf { Multi-parton Cross Sections at Hadron Colliders} }\\
\vspace*{3cm}
{\bf \large  Costas G. Papadopoulos$^1$ and
Ma\l gorzata~Worek$^{1,2}$}\\
\vspace{0.4cm}
{\it $^1$ Institute of Nuclear Physics, NCSR Demokritos,
15310 Athens, Greece\\
$^2$ Institute of Nuclear Physics Polish Academy of Sciences\\
 Radzikowskiego 152,
31-3420 Krakow, Poland\\ }
\vspace{3.5 cm}
{\bf Abstract}
\end{center}
\vspace{0.4 cm}

We present an alternative method to calculate cross sections for
multi-parton scattering processes in the Standard Model at leading
order.  The helicity amplitudes are computed using  recursion
relations in the number of particles, based on
Dyson-Schwinger equations whereas the summation over colour and helicity
configurations is performed by Monte Carlo methods. The computational
cost of our algorithm grows asymptotically as $3^n$, where $n$ is the
number of particles involved in the process, as opposed to the
$n!$-growth of the Feynman diagram approach.
Typical results for the total cross section, differential distributions
of invariant masses and transverse
momenta of partons are presented and cross checked by
explicit summation over colours.

\vspace{2 cm}
\begin{center}
\rule[.1in]{14.5cm}{.002in}

{\tt e-mail: costas.papadopoulos@cern.ch, malgorzata.worek@desy.de}
\end{center}

\setcounter{page}{1}
\pagestyle{plain}

\section{Introduction}

The simultaneous production of a large number of
energetic partons in high energy collisions of hadrons and leptons,
at the TeVatron or,
in future, at the LHC and the $e^{+}e^{-}$ Linear Collider,
 gives rise to events with many jets in the final
state. Quite often, these multi-jet events offer an important probe
of new physics as for example in the case of heavy particle decays in
the Standard Model and its extensions, {\it e.g.} the Minimal
Supersymmetric Standard Model. A particularly well known example
 is the  Higgs boson decay into four jets through $W/Z$ pairs.  Moreover,
the multi-jet events provide a significant background to the
discovery channels. The description of  these processes  is
 difficult even at leading
order,  because the corresponding amplitudes have to be
constructed from a very large number of Feynman
diagrams, making automation
the only  solution. As an example, in Tab. \ref{FD}.  the
number of Feynman diagrams relevant for the calculation of $gg \rightarrow
ng$  is collected. As can be seen, it grows asymptotically factorially
with the number of particles.

\begin{table}[h!]
\newcommand{\lstrut}{{$\strut\atop\strut$}}
\caption {\em  The number of Feynman diagrams
contributing to the total amplitude for $gg \rightarrow ng$.}
\vspace{0.2cm}
\label{FD}
\begin{center}
\begin{tabular}{l|l}
\hline \hline &  \\
$\textnormal{Process}$ & ~~~~~~
$\textnormal{N}_{\textnormal{\tiny FG}}$\\   &\\
\hline \hline  &\\
$gg \rightarrow   2g$ & 4\\
$gg \rightarrow   3g $ & 25  \\
$gg \rightarrow   4g $ & 220\\
$gg \rightarrow   5g $ & 2485\\
$gg \rightarrow   6g $ & 34300 \\
$gg \rightarrow   7g $ & 559405  \\
$gg \rightarrow   8g $ & 10525900  \\
$gg \rightarrow   9g $ & 224449225  \\
$gg \rightarrow   10g $ &5348843500  \\
& \\ \hline\hline
\end{tabular}
\end{center}
\end{table}

Another aspect of dealing with multi-particle amplitudes is the
systematic organisation of the summation over helicity
configurations and the $SU(N_c)$ colour algebra. If summation were
performed directly then $2^{n_{1}} \times 3^{n_{2}}$ helicity
configurations and  $8^{n_{g}}\times 3^{n_{q}} \times
3^{n_{\bar{q}}}$ colour configurations would have to be considered,
where $n_{g}$, $n_{q}$, $n_{\bar{q}}$ is the number of gluons,
quarks and antiquarks respectively while $n_{1}$ stands for the
number of fermions and massless bosons and $n_{2}$ is the number of
massive vector bosons.  Many of these configurations do not
contribute to the amplitude. However, it is very hard to predict
which should be kept in advance. The only way to achieve the
required efficiency is to use Monte Carlo techniques. A further
complication is connected to the integration over the
multi-dimensional phase space. In this report we present an approach
for efficient tree level calculations of matrix elements for
multi-parton final states which addresses the above problems and
therefore  improves the currently available
techniques\cite{Dobbs:2004qw,Maltoni:2002qb,Krauss:2001iv,Mangano:2002ea}.
In this paper, the development of a Monte Carlo summation over
colours for the full amplitude and of an efficient multi-particle
phase space integrator are presented. All algorithms have been
implemented in a Fortran 95 program that will be subsequently made
publicly available.

The layout of the paper is as follows. In Section \ref{sec1}  the
current status of available methods  is
briefly reviewed. Section \ref{sec2} describes the colour flow
decomposition.  Section \ref{sec3} presents the
recursive  relations and algorithms used to build the  amplitude. In
section \ref{sec4} we give the details of the internal organization
of our implementation of the algorithms. Here, we also introduce a
new approach to the colour structure evaluation. Its computational
complexity is  also briefly analyzed. In Section \ref{sec6},
numerical  results for the cross sections are presented together
with distributions  of the invariant mass and transverse
momentum.  The final section contains our summary  and an
outlook on future improvements  of the algorithm. An Appendix
contains a new algorithm for efficient  phase space point
generation.

\section{Dual amplitudes and colour decomposition}
\label{sec1}

For generality we consider $n$-gluon scattering
\begin{equation}
g(p_{1},\varepsilon_1,a_1) ~g(p_{2},\varepsilon_2,a_2)
\rightarrow g(p_{3},\varepsilon_3,a_3) \ldots g(p_{n},\varepsilon_n,a_n)
\end{equation}
with  external momenta $\{p_{i}\}_1^n$, helicities
$\{\varepsilon_{i}\}_1^n$ and colours $\{a_{i}\}_1^n$  of gluons
$i=1,\dots,n$ in the adjoint representation.
 The total amplitude can be expressed as a sum of single trace terms:

\begin{equation}
 {\cal
M}(\{p_{i}\}_{1}^{n},\{\varepsilon_{i}\}_{1}^{n},\{a_{i}\}_{1}^{n})
= \sum^{}_{I\in P(2,\ldots,n)} Tr(t^{a_1}t^{a_{\sigma_I(2)}}\ldots
t^{a_{\sigma_I(n)}}) {\cal A}_I
(\{p_{i}\}_{1}^{n},\{\varepsilon_{i}\}_{1}^{n})
\end{equation}
where $\sigma_I(2:n)$ represent the $I$-th permutation of the set
$\{2,\ldots,n\}$ and  $Tr(t^{a_1}t^{a_{\sigma_I(2)}}\ldots
t^{a_{\sigma_I(n)}})$ represents a
 trace of generators of the $SU(N_{c})$ gauge group in the fundamental
 representation.
  For processes involving quarks a similar expression
 can be derived \cite{Mangano:1990by}.  One of the most interesting aspects of this
 decomposition is the fact that  the ${\cal
 A}_I(\{p_{i}\}_{1}^{n},\{\varepsilon_{i}\}_{1}^{n})$ functions (called dual, partial or
 colour-ordered amplitudes), which
 contain all the kinematic information, are gauge invariant and cyclically
 symmetric in the momenta and helicities of gluons.   The
 colour ordered amplitudes are simpler than the full amplitude because
 they only receive contributions from diagrams  with a particular
 cyclic ordering of the external gluons (planar graphs).  For some processes up to six
 external partons simple and compact  expressions exist in
 literature \cite{Berends:1981rb,Parke:1985pn,Parke:1985ax,Kunszt:1985mg,Berends:1987cv,Mangano:1987xk,Mangano:1990by}.
 Moreover, for some special helicity combinations,
 short analytical forms, called the Parke-Taylor helicity amplitudes
 or Maximally
  Helicity Violating (MHV) amplitudes,  are known for  general
 $n$.  They were  first obtained by the Parke and  Taylor
\cite{Parke:1986gb} and later
 on proved to be correct in a recursive approach by Berends and
 Giele \cite{Berends:1987me}.
Recently, analytical expressions have also been obtained for other helicity
configurations \cite{Luo:2005rx,Britto:2005dg}, for an arbitrary number
of gluons. However, no analytic expressions for all helicity
configurations are known, and, with the exception of MHV
amplitudes, the known analytical expressions are usually cumbersome.
%
The simplicity of MHV amplitudes suggests that they can be used as
the basis of approximation schemes.
 Nevertheless, for large $n$, the  computation of scattering processes is
 still problematic and time consuming.  For example to  evaluate the
 full amplitude,  the $2^{n-1}\times(n-1)!$ configurations of colour ordered amplitudes,
 have to be considered,  where $2^{n-1}$ corresponds to the number of
 helicity configurations for massless particles. To obtain the cross
 section from the $n$-gluon amplitude one has  to square and sum
 over helicity and colour of the external gluons. The squared matrix
 element can be computed by
\begin{equation}
\sum_{\{a_i\}_{1}^{n}\{\varepsilon_{i} \}_{1}^{n}} |{\cal
M}(\{p_{i}\}_{1}^{n},\{\varepsilon_{i}
\}_{1}^{n},\{a_{i}\}_{1}^{n})|^2 = \sum_{\varepsilon}\sum_{IJ}{\cal
A}_{I}{\cal C}_{IJ}{\cal A}_{J}^{*} \label{colour_summation}
\end{equation}
where the $(n-1)!\times (n-1)!$ dimensional colour matrix can be written
in the most general form as follows:
\begin{equation}
{\cal C}_{IJ}= \sum_{1\ldots N_c}
Tr(t^{a_1}t^{a_{\sigma_I(2)}}\ldots t^{a_{\sigma_I(n)}})
Tr(t^{a_1}t^{a_{\sigma_J(2)}}\ldots t^{a_{\sigma_J(n)}})^{*}
\end{equation}

Needless to say that the evaluation of this matrix, is by its own a
formidable task in the standard approach.

An important step in the direction of simplification of these
calculations has already been taken by using helicity amplitudes
and a better organisation of the Feynman diagrams \cite{Kleiss:1985yh,Gunion:1985vc,
Xu:1986xb,Mangano:1987xk,Mangano:1990by,Berends:1987cv,Kuijf:1991kn}.
A significant
 simplification in these calculations has been made by introducing
 recursive relations \cite{Berends:1987me,Giele:1989vp}, which  express
 the $n$-parton currents in terms of  all currents up to $(n-1)$ partons.
  They are based on smaller building blocks which are
 just colour ordered vector and  spinor currents defined for the  off
mass shell particles.
 Extending the recursive approach beyond colour ordered
 amplitudes, {\it i.e.} to the full amplitude in any field theory,
 is possible.
Another approach in this direction
 based on the so called
 ALPHA algorithm \cite{Caravaglios:1995cd,Caravaglios:1998yr}
 or the Dyson-Schwinger
 recursion equations has been developed
 \cite{Argyres:1992cp,Argyres:1992js,Argyres:1992kt,Argyres:1992un,
 Argyres:1993wz,Argyres:2001sa,Draggiotis:1998gr,Kanaki:2000ey,Kanaki:2000ms,Draggiotis:2002hm},
 where the
 multi-parton amplitude can be  constructed without referring to
 individual Feynman diagrams.
 In the latter case, apart from the summation over
 colour in the colour flow basis, which will be briefly
described in the next section,
  integration over a continuous set of colour
 variables (as well as flavour) was introduced.
This integration technique, however, does not give a
 straightforward solution for the efficient merging of the parton
 level calculation with the parton shower evolution.

\section{The colour flow decomposition}
\label{sec2}

\begin{figure}[h!]
\begin{center}
\epsfig{file=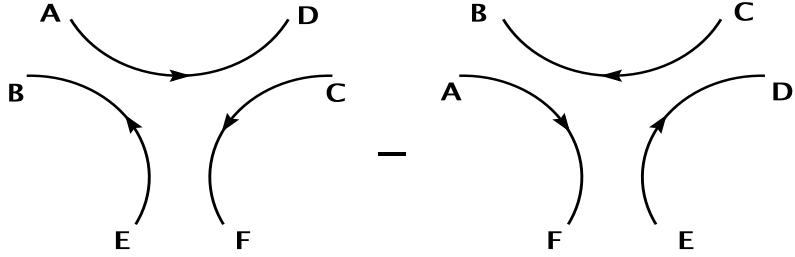,width=105mm,height=35mm}
\end{center}
\caption
{\it  Colour flows for the three gluon vertex.}
\label{colour_flow}
\end{figure}

First, let us briefly review
the colour approach used in  the original version
of {\tt HELAC} \cite{Kanaki:2000ey,Kanaki:2000ms}, a multipurpose
Monte Carlo generator for multi-particle final states
based on the Dyson-Schwinger recursion equations.
As was already mentioned, the colour connection or  colour flow
representation of the interaction vertices was used in this case.
This representation was introduced for the first
time in \cite{'tHooft:1973jz} and later studied in {\it e.g.}
\cite{Kanaki:2000ms,Maltoni:2002mq}.
The advantage of this color representation, as compared to the traditional one
is that the colour factors acquire a much simpler form, which moreover holds
for gluon as well as for quark amplitudes, leading to a unified approach
for any tree-order process involving any number of coloured partons.
Additionally, the usual
information on colour connections, needed by the parton shower Monte Carlo, is
automatically available, without any further calculation.
In this approach, the gluon field represented as
$A_{\mu}^{a}$ with $a=1,\ldots,N_{c}^{2}-1$
is treated as an $N_{c}\times N_{c}$ traceless matrix in colour
space \cite{Draggiotis:1998gr}. The new object
$(A_{\mu})_{AB}$  where $A,B=1,\ldots,N_{c}$ can be obtained by
multiplying each gluon field by the corresponding $t^{a}_{AB}$
matrix as follow:
\begin{equation}
(A_{\mu})_{AB}\equiv \sum_{a=1}^{N_c^2-1}t^{a}_{AB}A_{\mu}^{a}.
\end{equation}
The colour structure of the three gluon
vertex is given now by, see Fig.~\ref{colour_flow}.:
\begin{equation}
\sum_{a_i} f^{a_1a_2a_3}t_{AB}^{a_1}t_{CD}^{a_2}t_{EF}^{a_3}=
-\frac{i}{4}(\delta_{AD}\delta_{CF}\delta_{EB}-
\delta_{AF}\delta_{CB}\delta_{ED}) \label{3g-vertex}
\end{equation}
where, on the right hand side only products of $\delta$'s appear.
This colour structure shows how the colour flows
in the real physical process,
where gluons are represented by colour-anticolour
states in the colour space, and
reflects the fact that the colour remains unchanged on an
uninterrupted colour line.
\begin{figure}[h!]
\begin{center}
\epsfig{file=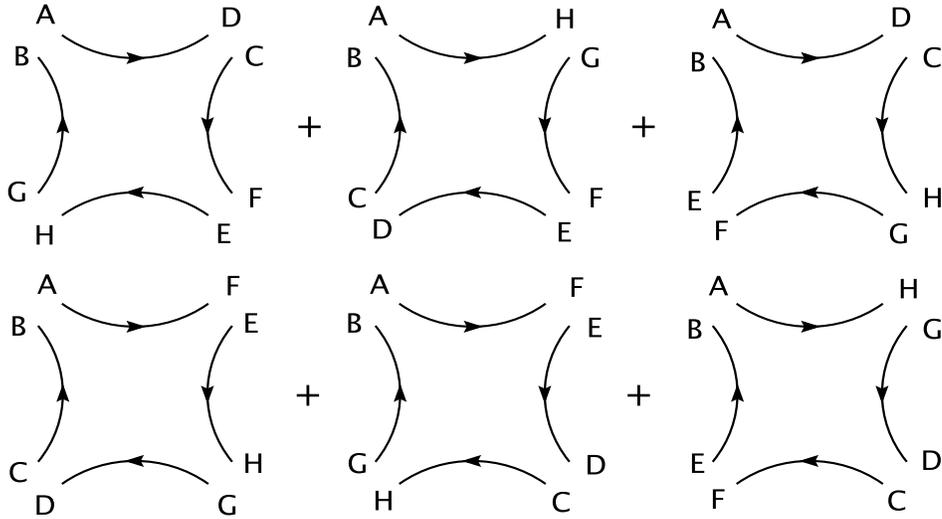,width=125mm,height=70mm}
\end{center}
\caption
{\it  Colour flows for the four gluon vertex.}
\label{colour_flow2}
\end{figure}
\begin{figure}[h!]
\begin{center}
\epsfig{file=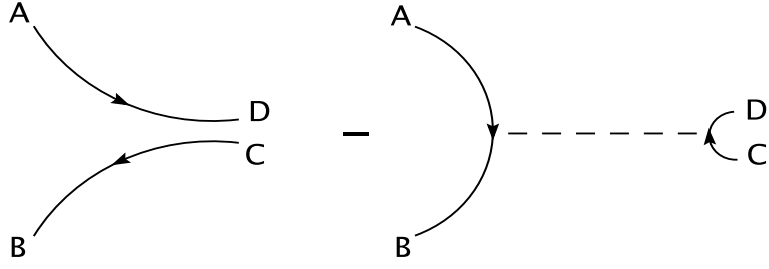,width=115mm,height=45mm}
\end{center}
\caption
{\it  Colour flows for the $q\bar{q}g$ vertex.}
\label{colour_flow3}
\end{figure}
For a four gluon vertex the following expression has to be considered:
\begin{equation}
\sum_{a_i}
 f^{a_1a_2x}f^{xa_3a_4}t^{a_1}_{AB}t^{a_2}_{CD}t^{a_3}_{EF}t^{a_4}_{GH}
\end{equation}
with three permutations of the $a_1a_2a_3$ indices, which correspond
to the six colour flows presented in Fig.\ref{colour_flow2}. Because
gluons have $N_c^{2}$ different colour states, described for
$N_c=3$ within the $U(3)$ group we have an additional unphysical
neutral $U(1)$ gluon. This neutral gluon does not couple to other
gluons as can be easily seen from Eq.(\ref{3g-vertex}). It couples
only to  quarks and acts as a colourless particle, see second part
of Eq.(\ref{qqg}). The  colour structure of the quark-antiquark-gluon
vertex can be described as follows, see Fig.\ref{colour_flow3}:
\begin{equation}
\sum_{a} t^{a}_{AB}t^{a}_{CD}=\frac{1}{2}(\delta_{AD}\delta_{CB}-
\frac{1}{N_c}\delta_{AB}\delta_{CD}). \label{qqg}
\end{equation}

Let us introduce a more compact notation and associate, to each gluon, a label
$(i,\sigma_i)$ which refers to the corresponding colour index of previous
equations, namely $1\rightarrow A$, $\sigma_1 \rightarrow B$ and so on.
The use of labels will be explained latter on.
With this notation the first term of the three gluon vertex is proportional
to:
\begin{equation}
\delta_{1\sigma_2}\delta_{2\sigma_3}\delta_{3\sigma_1}.
\end{equation}
For the same graph with inverted arrows a minus sign and
interchanged $2\leftrightarrow 2$ has to be included as well. The
momentum part of the vertex, $V^{\mu_1\mu_2\mu_3}$ is still the
usual one and in our notation simply given by:
\begin{equation}
g^{12}(p_1-p_2)^3+g^{23}(p_2-p_3)^1+g^{31}(p_3-p_1)^2.
\end{equation}
For the $q\bar{q}g$ vertex we associate a label $(i,0)$ for quark
and $(0,\sigma_i)$ for antiquark. Finally the four gluon vertex
 is given by a colour factor proportional to:
\begin{equation}
\delta_{1\sigma_3}\delta_{3\sigma_2}\delta_{2\sigma_4}\delta_{4\sigma_1}
\end{equation}
with six possible permutations, and a Lorentz part,
$G^{\mu_1\mu_2\mu_3\mu_4}$,
\begin{equation}
2g^{13}g^{24}-g^{12}g^{34}-g^{14}g^{23}
\end{equation}
where all three permutations should be included.

To make use of the colour representation described so far, let us assign,
to each external gluon, a label $(i,\sigma_{I}(i))$, to a quark $(i,0)$
and to antiquark $(0,\sigma_{I}(i)))$, where $i=1\ldots n$ and
$\sigma_I(i), I=1\ldots n!$ being a permutation of $\{1\ldots n\}$.
Since all elementary colour factors appearing in the colour decomposition
of the vertices are proportional to $\delta$ functions
the total colour factor can be given by

\begin{equation}
{\cal
M}(\{p_{i}\}_{1}^{n},\{\varepsilon_{i}\}_{1}^{n},\{c_i,a_i\}_{1}^{n})=
\sum^{}_{I=P(2,\ldots,n)} D_I \;\; {\cal A}_I
(\{p_{i}\}_{1}^{n},\{\varepsilon_{i}\}_{1}^{n})
\end{equation}

\begin{eqnarray}
{\cal D}_I=\delta_{1\sigma_I(1)}\delta_{2\sigma_I(2)}
\ldots\delta_{n\sigma_I(n)}\,,
\end{eqnarray}
\noindent
The colour matrix  defined as
\begin{equation}
{\cal C}_{IJ} = \sum_{IJ} {\cal D}_I {\cal D}_J^\dagger
\end{equation}
with the summation running over all colours, $1,\dots,N_c$
has a very simple representation now
\begin{equation}
{\cal C}_{IJ} = N_c^{m(\sigma_I,\sigma_J)}
\end{equation}
where $1\le m(\sigma_I,\sigma_J)\le n$ counts how many common cycles
the permutations $\sigma_I$ and $\sigma_J$ have. The practical
implementation of these ideas is straightforward. Given the
information on the external particles contributing to the process we
associate colour labels of the form $(i,\sigma_i)$ depending on
their flavour. According to the Feynman rules the higher level
sub-amplitudes are built up. Summing over all $n!$ colour connection
configurations, where $n$ is the number of gluons and $q\bar{q}$
pairs in the process using the colour matrix ${\cal C}_{IJ}$ we get
the total squared amplitude. It is worthwhile to note that summing
over all colour configurations is efficient as long as the number of
particles is smaller than ${\cal O}(8)$. If the number of gluons
and/or
 $q\bar{q}$ pairs  is higher
than ${\cal O}(8)$, since the number of colour flows in general
grows like $(n-1)!$, it also starts to be problematic from the
computational point of view. For multi-colour processes other
approaches have to be considered. The natural solution would be to
replace the summation over all colour connections by a Monte Carlo.
However, Monte Carlo summation
 is not straightforward  in the color flow approach because of
the destructive interferences between different colour flows that
can give a negative contribution to the squared matrix element.

In the limit $N_c\to \infty$, only the diagonal terms, ${I}={J}$, survive,
and all ${\cal O}(N_{c}^{-2})$ terms can be safely neglected
both in the colour matrix and in the $|{\cal A}_{I}|^{2}$. The interferences
between different colour flows vanish in this limit.
In the so called Leading Colour Approximation (LCA), the squared amplitude
for the purely gluonic case is given by
\begin{equation}
\sum_{a,\varepsilon} |{\cal
M}(\{p_{i}\}_{1}^{n},\{\varepsilon_{i}\}_{1}^{n},\{a_{i}\}_{1}^{n})|^2
=N_c^{n-2}(N_c^{2}-1)\sum_{\varepsilon}\sum_{I}|{\cal A}_{I}|^{2}.
\end{equation}
The term $N_c^{2}-1$ instead of $N_c^2$, which of course are
equivalent in the limit $N_c\to \infty$, has been kept in order to
reproduce the exact results for $n=4$ and $n=5$. In case when
$q\bar{q}$ pairs are present the colour factor
$N_c^{n-2}(N_c^{2}-1)$ is still  the same, however
$n=n_g+n_{q\bar{q}}$ in this case, where $n_g$, $n_{q\bar{q}}$ is
the number of gluons and $q\bar{q}$ pairs respectively. This
simplification of the colour matrix speeds up the calculation.
Moreover, Monte Carlo summation over colours can now be performed.

\section{Dyson-Schwinger recursion relations }
\label{sec3}

\begin{figure}[h!]
\begin{center}
\epsfig{file=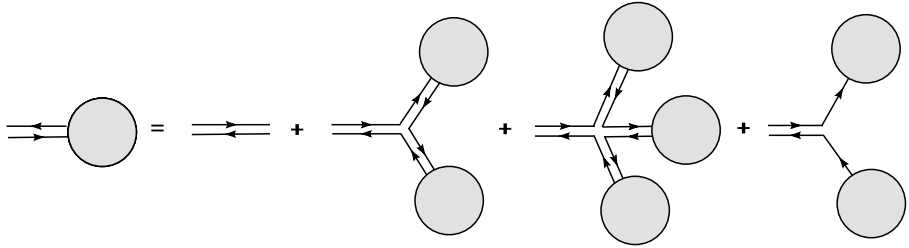,width=120mm}
\end{center}
\caption
{\it Recursive equation for an off mass shell gluon of momentum P. }
\label{ds_gluon}
\end{figure}

Let us now review
 the recursive relations based on Dyson-Schwinger
equations for the calculation of  partial amplitudes
\cite{Draggiotis:2002hm}.
These equations give recursively the $n-$point Green's functions
in terms of the $1-$, $2-$,$\ldots$, $(n-1)-$point functions. They
hold all the information for the fields and their interactions for any
number of external legs and to all orders in perturbation theory.
We will concentrate here on the gluon, quark and antiquark
recursion relations, however,
in the same way recursive equations for leptons and gauge bosons
can be obtained. The
diagrammatic picture  behind the recursive relations is actually quite
simple. The tree-level recursive equation can be
diagrammatically presented as shown in Fig.\ref{ds_gluon}.
Let $p_{1},p_{2},\ldots,p_{n}$
represent the external momenta involved in the scattering process taken
to be incoming.
In order to write down the recursive relation explicitly, we first
define a set of four vectors
$[A^{\mu}(P);(A,B)]$, which describes any sub-amplitudes
from which a gluon with momentum $P$ and colour-anticolour assignment $A,B$
can be constructed.
The momentum $P$ is given as a sum of external particles momenta.
Accordingly we define a set of four-dimensional spinors
$[\psi(P);(A,0)]$ describing any sub-amplitude
from which a quark with momentum P  and colour $A$ can be constructed and by
$[\bar{\psi}(P);(0,B)]$  a set of four-dimensional antispinors for antiquark with
 anticolour $B$.
The Dyson-Schwinger recursion equation for a gluon can
be written as follows:
\begin{equation}
[A^{\mu}(P);(A,B)]=
\sum_{i=1}^{n}[~\delta_{P=p_{i}}~A^{\mu}(p_{i});(A,B)_i]
\end{equation}
\[
+ \sum_{P=p_{1}+p_{2}}
[~(ig)  ~\Pi^{\mu}_{\rho} ~V^{\rho\nu\lambda}(P,p_1,p_2)A_{\nu}(p_{1})A_{\lambda}(p_{2})\sigma(p_1,p_2);(A,B)=(C,D)_1\otimes(E,F)_2]
\]
\[
-\sum_{P=p_{1}+p_{2}+p_{3}}
[~(g^2)~\Pi^{\mu}_{\sigma}~ G^{\sigma\nu\lambda\rho}(P,p_1,p_2,p_3)
 A_{\nu}(p_{1})
A_{\lambda}(p_{2})A_{\rho}(p_{3})\sigma(p_1,p_2+p_3);
\]
\[
(A,B)=(C,D)_1\otimes(E,F)_2\otimes(G,H)_3]
\]
\[
+ \sum_{P=p_{1}+p_{2}}[~(ig) ~\Pi^{\mu}_{\nu} ~\bar{\psi}
(p_1)\gamma^{\nu}\psi(p_2)\sigma(p_1,p_2);(A,B)=(0,D)_1\otimes(C,0)_2]
\]
where $A,B,C,D,E,F,G,H=1,2,3$.
The rules for  merging colour and anticolour of the
particles will be explained in the next section.
The $V^{\mu\nu\lambda}(P,p_1,p_2)$ and
$G^{\mu\nu\lambda\rho}(P,p_1,p_2,p_3)$ functions  are the three- and four-
gluon vertices presented in the previous sections
and the symbol $\sigma(p_1,p_2)$ is the sign function
which takes into account the Fermi sign when two identical fermions are
interchanged. The exact form of this function
can be found in Ref.\cite{Kanaki:2000ey,Draggiotis:2002hm}.
The  sums are over all
combinations of $p_1,p_2$ or $p_1,p_2,p_3$ that sum up to $P$.
The propagator of the gluon is given by:
\begin{equation}
\Pi_{\mu\nu}=\frac{-ig_{\mu\nu}}{P^2}.
\end{equation}
\begin{figure}[h!]
\begin{center}
\epsfig{file=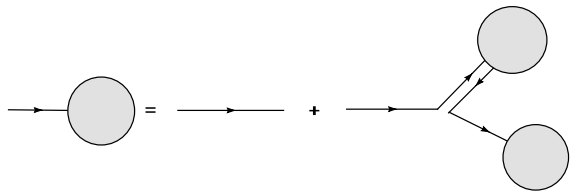,width=75mm}
\end{center}
\caption
{\it Recursive equation for an off mass shell quark of momentum P. }
\label{ds_quark}
\end{figure}
\begin{figure}[h!]
\begin{center}
\epsfig{file=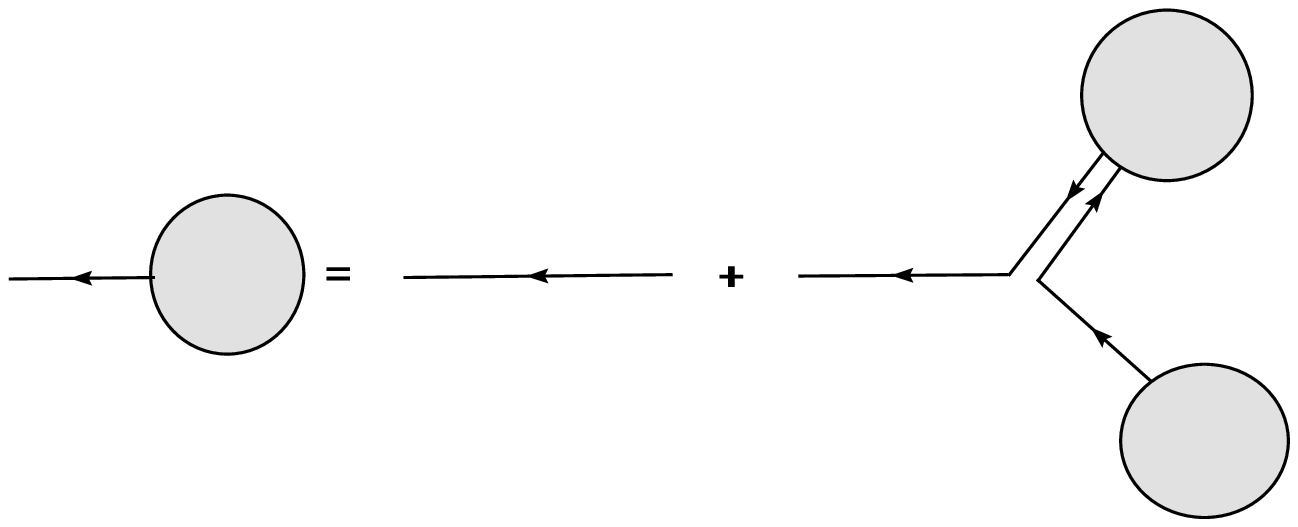,width=75mm}
\end{center}
\caption
{\it Recursive equation for an off mass shell antiquark of momentum P. }
\label{ds_antiquark}
\end{figure}
For a quark of momentum $P$ we have, Fig.\ref{ds_quark}:
\begin{equation}
\label{quark_ds}
[\psi(P);(A,0)]=\sum_{i=1}^{n}[~\delta_{P=p_{i}}~\psi(p_i);(A,0)_i]
\end{equation}
\[
+ \sum_{P=p_1+p_2}[~(ig)~{\cal P} ~A^{\mu}(p_1)\gamma_{\mu}\psi(p_2)
\sigma(p_1,p_2); (A,0)=(B,C)_1\otimes(D,0)_2]
\]
where ${\cal P}$ is the propagator
\begin{equation}
{\cal P}=\frac{iP\!\!\!/}{P^2}.
\end{equation}
Finally for an antiquark, Fig.\ref{ds_antiquark}:
\begin{equation}
[\bar{\psi}(P);(0,A)]=\sum_{i=1}^{n}[~\delta_{P=p_{i}}~\bar{\psi}(p_i);(0,A)_i]
\end{equation}
\[
+ \sum_{P=p_1+p_2}[~(ig)
~\bar{\psi}(p_2)A^{\mu}(p_1)\gamma_{\mu}~\bar{\cal P}\sigma(p_1,p_2)
;(0,A)=(B,C)_1\otimes(0,D)_2]
\]
where
\begin{equation}
\bar{\cal P}=\frac{-iP\!\!\!/}{P^2}.
\end{equation}
In the same spirit, the recursion equations for all leptons and gauge
bosons can be written down. With this algorithm we can thus compute
the scattering amplitude  for any initial and final states taking
into account particle masses as well.  In particular any colour
structure can be assigned to the external legs. Finally this
approach has an exponential growth of computational time with the
number of external particles instead of the factorial growth when
Feynman graphs are considered. It  can be farther
optimised in order to reduce computational complexity by replacing
each four gluon vertex by a three particle vertex by introducing the
auxiliary field represented by the antisymmetric tensor
$H^{a}_{\mu\nu}$. This new field has a quadratic term without
derivatives and, therefore, has no independent dynamics. The part of the QCD
Lagrangian that describes the four gluon vertex
\begin{equation}
\label{lag1}
{\cal L}=-\frac{1}{4}F_{\mu\nu}^{a}F^{\mu\nu a},
~~~~~~~~~~~F^a_{\mu\nu}=
\partial_{\mu}A_{\nu}^{a}-\partial_{\nu}A_{\mu}^{a}+gf^{abc}A_\mu^{b}
A_{\nu}^{c}
\end{equation}
can be rewritten in terms of the auxiliary field as follows:
\begin{equation}
{\cal L}=-\frac{1}{2}H_{\mu\nu}^{a}H^{\mu\nu a}+\frac{1}{4}H_{\mu\nu}^{a}
F^{\mu\nu a}.
\end{equation}
A single interaction term of the form
$H^{\mu\nu a}A^{b}_{\mu}A^{c}_{\nu}$
is left instead of interaction terms represented by Eq.(\ref{lag1}).
The recursion for the gluons changes slightly,
in fact only for the four-gluon vertex part. However,
we have an additional equation for the auxiliary field:
\begin{equation}
\label{aux1} [A^{\mu}(P);(A,B)]=
\sum_{i=1}^{n}[~\delta_{P=p_{i}}~A^{\mu}(p_{i});(A,B)_i]
\end{equation}
\[
+\sum_{P=p_{1}+p_{2}}
[~(ig)~\Pi^{\mu}_{\rho}~
V^{\rho\nu\lambda} (P,p_1,p_2) A_{\nu}(p_{1})A_{\lambda}(p_{2})
\sigma(p_1,p_2);(A,B)=(C,D)_1\otimes(E,F)_2]
\]
\[
+\sum_{P=p_{1}+p_{2}}[~(ig)~\Pi^{\mu}_{\sigma}~(g^{\sigma\lambda}
g^{\nu\rho}-g^{\nu\lambda}g^{\sigma\rho})~A_{\nu}(p_1)H_{\lambda\rho}(p_2)
\sigma(p_1,p_2);(A,B)=(C,D)_1\otimes(E,F)_2]
\]
\[
+\sum_{P=p_{1}+p_{2}}[~(ig) ~\Pi^{\mu}_{\nu}~\bar{\psi}
(p_1)\gamma^{\nu}\psi(p_2)\sigma(p_1,p_2);(A,B)=(0,D)_1\otimes(C,0)_2]
\]
and
\begin{equation}
\label{aux2}
[H^{\mu\nu}(P);(A,B)]= \sum_{P=p_{1}+p_{2}} [~(ig) ~(g^{\mu\lambda}
g^{\nu\rho}-g^{\nu\lambda}g^{\mu\rho}) ~A_{\lambda}(p_1)A_{\rho}(p_2)
\sigma(p_1,p_2);
\end{equation}
\[
(A,B)=(C,D)_1\otimes(E,F)_2].
\]
A few  comments are now in order. First, all calculations were performed
in the light cone representation and
all momenta were taken to be incoming. Second,
this new $H_{\mu\nu}$ field has six components. Additionally, as we can
see in Eq.(\ref{aux1}) and Eq.(\ref{aux2}) the colour structure of this
new vertices, remains the same as in case of the three gluon vertex.
The number of the types of the sub-amplitudes one has to
calculate  is in general doubled, however their structure is much
simpler, which saves computational time while the iteration steps are performed.

After $n-1$ steps, where
$n$ is the number of particles under consideration,
one can get the total amplitude.
The scattering amplitude can be calculated by any of the following
relations, depending on the process under consideration,
\begin{equation}
{\cal A}(\{p_{i}\}_{1}^{n},\{\varepsilon_{i}\}_{1}^{n})=
\left\{
\begin{array}{ll}
\hat{A}^\mu(P_i) A_\mu(p_i) & \mbox{where $i$ corresponds to gluon}
\\ \hat{\psb}(P_i){\psi}(p_i) & \mbox{where $i$  corresponds to
quark line}  \\ \psb(p_i)\hat{\psi}(P_i) & \mbox{where
$i$  corresponds to  antiquark line}  \\
\end{array}
\right.
\end{equation}
where
\[ P_i=\sum_{j\not= i}p_j,\]
so that $P_i+p_i=0$. The functions with  hat are given by the
previous expressions except for the propagator term which is removed
by the amputation procedure. This is because the
outgoing momentum $P_{i}$ must be on shell.
The initial conditions are given by
\bqa
A^\mu(p_i)&=&\epsilon^\mu_\lambda(p_i) ,    \lambda=\pm1,0
\nn\\
\psi(p_i)&=&\left\{
\begin{array}{ll}
u_\lambda(p_i)
&\mbox{if $p_i^0\geq0$} \\
v_\lambda(-p_i)
&\mbox{if $p_i^0\leq0$} \\
\end{array}
\right.
\nn\\
\psb(p_i)&=&\left\{
\begin{array}{ll}
\bar{u}_\lambda(p_i)
&\mbox{if $p_i^0\geq0$} \\
\bar{v}_\lambda(-p_i)
&\mbox{if $p_i^0\leq0$} \\
\end{array}
\right.
\label{ampl}\eqa
where the explicit form of $\epsilon^\mu_\lambda,u_\lambda,
v_\lambda,\bar{u}_\lambda,\bar{v}_\lambda$
is given in the Ref.\cite{Kanaki:2000ey}.

\section{Organisation of the calculation}
\label{sec4}

The recursive algorithm presented in the previous section has the
advantage  that any colour representation can be used in order to
assign colour degrees of freedom to the external legs in the process
under consideration. It can be used either for colour ordered
amplitudes  or for the full amplitudes as well. The latter case is in
fact the topic of this section. Moreover, an alternative method for
taking into account the colour structure of scattering partons,
based on regular colour configuration assignments as compared to
colour flow ones, will be introduced. First, however, the
organisation of the calculation will be explained briefly in order
to better describe the general structure. Contrary to the original
{\tt HELAC} \cite{Kanaki:2000ey,Kanaki:2000ms} approach, in this new
version  the computational part consists of one phase only.  This is
not optimal for electroweak processes with moderate number of
external particles, but it becomes quite efficient for processes
with many particles, especially when only a few  species of
particles are involved (scalar amplitudes, gluon amplitudes in QCD,
etc).  The vertices described by the Standard Model Lagrangian are
implemented in fusion rules that dictate the way the subamplitudes,
at each level of the recursion relation, will be merged, in order to
produce a higher level subamplitude. In case when a quark is combined with
an antiquark, for example, there are three possible
subamplitudes which describe three different  intermediate states
incorporated inside the fusion rules, namely $\gamma$, $Z$ and $g$.

Let us now present  the colour merging rules which are evaluated
iteratively at the subamplitude level. During each iteration
when two particles are combined  their corresponding colour
assignments  are combined.  We have three possibilities to obtain a gluon $(A,B)$
described by the recursive relation Eq.(\ref{aux1}). First,  it can be
obtained when a quark with colour
assignment $(C,0)$ is merged with an  antiquark of anticolour
$(0,D)$. Second, when two gluons $(C,D)$ and $(E,F)$  are combined,
where $A,B,C,D,E,F=1,2,3$, and finally when a gluon and an
auxiliary field $H$ are combined. In the last case, the colour structure
of the vertex is identical to the three-gluon vertex. So we end up
with the following rules:
\begin{eqnarray}
&&(A,B)\leftarrow (C,0~)\otimes(~0,D),\\
&&(A,B)\leftarrow (C,D)\otimes(E,F).
\end{eqnarray}
In the first case the merging occurs according to the rule presented
in Eq.(\ref{qqg}). and the following gluon can be produced:
\begin{eqnarray}
&&(A,B)=(C,0)\otimes(0,D)=(C,D), ~~~~~~ {\textnormal{if}} ~C\ne D.
\end{eqnarray}
However, the situation is more complex when a quark and an antiquark have
the same  colour and anticolour:
\begin{eqnarray}
&&(A,B)=(C,0)\otimes(0,D)=(1,1)_{w_1}\oplus(2,2)_{w_2}\oplus(3,3)_{w_3},
~~~~~~ {\textnormal{if}} ~C=D.
\end{eqnarray}
with $\sum w_i=0$. We thus have three possibilities with different
weights, which for instance are $w_1=w_2=-1/3$, $w_3=+2/3$ in case
$C=D=3$.

In the second case, when gluons are combined
according to Eq.(\ref{3g-vertex}) we have the following options:
\begin{eqnarray}
&&(A,B)=(C,D)\otimes(E,F)=(E,D), ~~~~~~ {\textnormal{if}} ~C=F ~~~~~ {\textnormal{and}} ~E\ne D\\
&&(A,B)=(C,D)\otimes(E,F)=(C,F), ~~~~~~ {\textnormal{if}} ~D=E
 ~~~~~ {\textnormal{and}} ~ C\ne F.
\end{eqnarray}
Moreover, we have an additional possibility when colours and
anticolours of gluons are the same. One gets, in this case, a gluon
in two colour states with different weight:
\begin{equation}
(A,B)=(C,D)\otimes(E,F)=(C,C)_{+1}\oplus(D,D)_{-1}, ~~~
{\textnormal{if}} ~~~ C=F ~~~ E=D ~~~ {\textnormal{and}} ~~~ C\ne D.
\end{equation}
Finally for $C=D=E=F$ the result vanishes identically.

To obtain the quark, $(A,0)$, described by the recursive relation
 Eq.(\ref{quark_ds}) we have to combine a gluon with another quark
one more time according to the rule presented in Eq.(\ref{qqg}):
\begin{equation}
 (A,0)=(B,C)\otimes(D,0).
\end{equation}
We have two possibilities for colour assignment:
\begin{eqnarray}
&&(A,0)=(B,C)\otimes(D,0)=(B,0), ~~~~~~ {\textnormal{if}} ~C=D,\\
&&(A,0)=(B,C)\otimes(D,0)=(D,0), ~~~~~~ {\textnormal{if}} ~B=C.
\end{eqnarray}
For the antiquark the situation is the same, so we will not
elaborate it here.

For the $q\bar{q}$ interactions with $\gamma,Z^{0},W^{\pm}$ or Higgs boson
the situation is very simple and we have only one possibility:
\begin{equation}
(A,0)\otimes (0,B)=(0,0)~~~~~~~~~~~~~~~\textnormal{if} ~A=B.
\end{equation}

The sum over colour can be performed in this way by considering all
possible colour-anticolour configurations according to the above
rules. However, the procedure can be facilitated by Monte Carlo
methods where a particular colour-anticolour configuration is
randomly selected. As we will see an important gain in computational
efficiency is achieved within this framework.

The {\it necessary} condition which
must be fulfilled, while the particular colour assignment for the external
coloured particles is chosen, is that the number
of colour and anticolour of each type, is the same. Otherwise
the particles can not be connected by colour flow lines and the
amplitude is identically zero. In the Monte Carlo over colours method
one has to multiply the squared matrix element
by a coefficient that counts the number of non zero colour
configurations.

The number of non zero colour configurations, according to the above mentioned
necessary condition is given by
\begin{equation}
N_{\textnormal{\tiny CC}}
=\sum_{A=0}^{n_{q}}~\sum_{B=0}^{n_{q}-A}~\sum_{C=0}^{n_{q}-A-B}
\bigg( \frac{n_{q}!}{A!B!C!} \bigg)^2 ~\delta(n_{q}=A+B+C)
\label{ncc}
\end{equation}
where $n_{q}$ is the total number of colours and $A,B,C$ are the
numbers of the colour type $1$, $2$ and $3$ respectively.
As we already stated, the condition
given by Eq.(\ref{ncc}) is necessary but not sufficient. Among this
set there are still configurations  which do not give contributions
to the total amplitude.
\begin{table}[h!]
\newcommand{\lstrut}{{$\strut\atop\strut$}}
\caption {\em  The number of colour configurations
for the processes with gluons only.
$N_{\textnormal{{\tiny CC}}}^{\textnormal{{\tiny ALL}}}$
corresponds to all possible colour configurations, while
$N_{\textnormal{{\tiny CC}}}$ corresponds to the number of colour
configurations calculated using formula
Eq.(\ref{ncc}). In the third column the ratio
$\textnormal{N}_{\textnormal{\tiny CC}}
/\textnormal{N}^{\textnormal{\tiny ALL}}_{\textnormal{\tiny CC}}$
is presented. In the last column the number of
non vanishing colour configurations evaluated by MC
$\textnormal{N}_{\textnormal{\tiny CC}}^{\textnormal{\tiny F}}$
(in percentage) inside the
$\textnormal{N}_{\textnormal{\tiny CC}}$ is shown.  }
\vspace{0.2cm}
\label{numbers}
\begin{center}
\begin{tabular}{l|l|l|l|l}
\hline \hline & &&& \\
$\textnormal{Process}$ &
~~~~$\textnormal{N}^{\textnormal{\tiny ALL}}_{\textnormal{\tiny CC}}$&
~~~~$\textnormal{N}_{\textnormal{\tiny CC}}$
&$\textnormal{N}_{\textnormal{\tiny CC}}
/\textnormal{N}^{\textnormal{\tiny ALL}}_{\textnormal{\tiny CC}}$
& ~~$\textnormal{N}_{\textnormal{\tiny CC}}^{\textnormal{\tiny F}}$
($\%$)\\   &&&&\\
\hline \hline & &&&
\\
$gg \rightarrow   2g$ & 6561       & 639       &~~0.0974 &~~~~59.1\\
$gg \rightarrow   3g$ & 59049      & 4653      &~~0.0788 &~~~~68.4  \\
$gg \rightarrow   4g$ & 531441     & 35169     &~~0.0662 &~~~~77.4  \\
$gg \rightarrow   5g$ & 4782969    & 272835    &~~0.0570 &~~~~85.0  \\
$gg \rightarrow   6g$ & 43046721   & 2157759   &~~0.0501 &~~~~90.4\\
$gg \rightarrow   7g$ & 387420489  & 17319837  &~~0.0447 &~~~~94.0 \\
$gg \rightarrow   8g$ & 3486784401 & 140668065 &~~0.0403 &~~~~96.4 \\
& & &&\\ \hline\hline
\end{tabular}
\end{center}
\end{table}
\begin{table}[h!]
\newcommand{\lstrut}{{$\strut\atop\strut$}}
\caption {\em  The number of colour configurations
for the processes with gluons and one or two $q\bar{q}$ pairs.
$N_{\textnormal{{\tiny CC}}}^{\textnormal{{\tiny ALL}}}$
corresponds to all possible colour configurations, while
$N_{\textnormal{{\tiny CC}}}$ corresponds to the number of colour
configurations calculated using formula
Eq.(\ref{ncc}). In the third column the ratio
$N_{\textnormal{{\tiny CC}}}/
N_{\textnormal{{\tiny CC}}}^{\textnormal{{\tiny ALL}}}$ is presented.
In the last column the number of
non vanishing colour configurations evaluated by MC
$\textnormal{N}_{\textnormal{\tiny CC}}^{\textnormal{\tiny F}}$
(in percentage) inside the
$\textnormal{N}_{\textnormal{\tiny CC}}$ is shown.}
\vspace{0.2cm}
\label{numbers2}
\begin{center}
\begin{tabular}{l|l|l|l|l}
\hline \hline & && &\\
$\textnormal{Process}$ & ~~~
$\textnormal{N}^{\textnormal{\tiny ALL}}_{\textnormal{\tiny CC}}$&
~~~~$\textnormal{N}_{\textnormal{\tiny CC}}$ &
$\textnormal{N}_{\textnormal{\tiny CC}}/
\textnormal{N}^{\textnormal{\tiny ALL}}_{\textnormal{\tiny CC}}$
& ~~$\textnormal{N}_{\textnormal{\tiny CC}}^{\textnormal{\tiny F}}$
($\%$)\\   & &&&\\
\hline \hline &&  &&\\
$gg \rightarrow  u\bar{u}$    &729       &93       &~~0.1276 &~~~~93.5\\
$gg \rightarrow  gu\bar{u}$   &6561      &639      &~~0.0974&~~~~91.6\\
$gg \rightarrow  2gu\bar{u}$  &59049     &4653     &~~0.0788&~~~~92.6 \\
$gg \rightarrow  3gu\bar{u}$  &531441    &35169    &~~0.0662&~~~~94.6  \\
$gg \rightarrow  4gu\bar{u}$  &4782969   &272835   &~~0.0570&~~~~96.4  \\
$gg \rightarrow  5gu\bar{u}$  &43046721  &2157759  &~~0.0501&~~~~97.8\\
$gg \rightarrow  6gu\bar{u}$  &387420489 &17319837 &~~0.0447&~~~~98.6 \\
& && &\\ \hline&& & & \\
$gg \rightarrow  c\bar{c}c\bar{c}$    &6561     &639     &~~0.0974&~~~~99.1\\
$gg \rightarrow  gc\bar{c}c\bar{c}$   &59049    &4653    &~~0.0788&~~~~98.8\\
$gg \rightarrow  2g c\bar{c}c\bar{c}$ &531441   &35169   &~~0.0662&~~~~99.0\\
$gg \rightarrow  3g c\bar{c}c\bar{c}$ &4782969  &272835  &~~0.0570&~~~~99.3\\
$gg \rightarrow  4g c\bar{c}c\bar{c}$ &43046721 &2157759 &~~0.0501&~~~~99.6\\
& && &\\ \hline\hline
\end{tabular}
\end{center}
\end{table}

In Tab.~\ref{numbers}. and Tab.~\ref{numbers2}.
different numbers of colour configurations in the process
under consideration are presented. In the first column
the total number of all colour configurations is listed, where
$\textnormal{N}^{\textnormal{\tiny ALL}}_{\textnormal{\tiny CC}}
=3^{n_{q}+n_{\bar{q}}}$ and  $n_{q}$,
$n_{\bar{q}}$ are the number of quarks and antiquarks respectively
and gluons are treated as $q\bar{q}$ pairs. The second column represents
results for the number of non vanishing
colour configurations $\textnormal{N}_{\textnormal{\tiny CC}}$
 calculated using
Eq.(\ref{ncc}). In the next column the ratio $N_{\textnormal{{\tiny CC}}}/
N_{\textnormal{{\tiny CC}}}^{\textnormal{{\tiny ALL}}}$ is presented.
The number of colour configurations (in percentage) inside the
$\textnormal{N}_{\textnormal{\tiny CC}}$ set which finally
gives rise to non zero amplitudes
is shown in the last column. Those numbers
$\textnormal{N}_{\textnormal{\tiny CC}}^{\textnormal{\tiny F}}$ are
evaluated by Monte Carlo.
Note that while the number of external particles is increased
the corresponding number of vanishing colour configurations in the third column
is decreased, as we can see in Tab.~\ref{numbers}.

As far as the summation over the helicity configurations is concerned
there are two possibilities, either the explicit summation over all
helicity configurations or a Monte Carlo approach. In the latter case,
for example for gluon, it is achieved  by introducing
the polarisation vector
\begin{equation}
\varepsilon^{\mu}_{\phi}(p)=e^{i\phi}\varepsilon^{\mu}_{+}(p)+
e^{-i\phi}\varepsilon^{\mu}_{-}(p),
\end{equation}
where $\phi \in (0,2\pi)$. By integrating over
$\phi$ we can obtain the sum over helicities
\[
\frac{1}{2\pi}\int_{0}^{2\pi}d\phi ~\varepsilon^{\mu}_{\phi}(p)
(\varepsilon^{\nu}_{\phi}(p))^{*}=\sum_{\lambda=\pm}
\varepsilon_{\lambda}^{\mu}(p)(\varepsilon_{\lambda}^{\nu}(p))^{*}.
\]

To determine the cost of computation of the $n-$point amplitude
using the algorithm based on Dyson-Schwinger equations, one
has to count the number of operations \cite{Petros-thesis}.
There are $\left( \begin{array}{l} n \\ k \end{array}\right)$
momenta at each level. The total amount of sub-amplitudes corresponding
to those momenta is  simply given by:
\begin{equation}
\sum_{k=1}^{n-1}
\left(
\begin{array}{l}
n\\k
\end{array}
\right)=2^n-2
\end{equation}
where $n$ is the number of particles involved in the calculation.
Moreover  one has to count how many ways exist to split a number of level $k$
to two numbers of levels $k_1$ and $k_2$.
The last step is to sum over all levels.
The total number of operations that should
be performed in the case when only three-point vertices exist is then
\begin{equation}
\sum_{k=1}^{n-1}
\left(
\begin{array}{l}
n\\k
\end{array}
\right)
\sum_{l=1}^{k-1}
\left(
\begin{array}{l}
k\\l
\end{array}
\right)
=\sum_{k=1}^{n-1}
\left(
\begin{array}{l}
n\\k
\end{array}
\right)
\{2^k-2\}=3^n-3\cdot2^n+3.
\end{equation}
In the limit $n\rightarrow \infty$ the number of operations grows like $3^n$
instead of the $n!$ growth in the Feynman graph approach.
When the Monte Carlo over colour structures is performed and only
one particular colour configuration is randomly chosen the
computational cost of this algorithm is given exactly by this expression.
Otherwise, this formula must be
multiplied by the number of non zero colour configurations for the process
under consideration.

\section{Numerical Results}
\label{sec6}
\begin{table}[h!]
\newcommand{\lstrut}{{$\strut\atop\strut$}}
\caption {\em  Results for the total cross section
for processes with gluons and quarks with up to two $q\bar{q}$ pairs.
$\sigma_{\textnormal{{\tiny EXACT}}}$ corresponds to summation over
all possible colour configurations, while
$\sigma_{\textnormal{{\tiny MC}}}$ corresponds to
 Monte Carlo summation.}
\vspace{0.2cm}
\label{tab1}
\begin{center}
\begin{tabular}{l|l|l|l|l}\hline \hline
&&&&\\
~~$\textnormal{Process}$ & ~~~~~~~$\sigma_{\textnormal{\tiny EXACT}}$
$\pm$ $\varepsilon$ $\textnormal{(nb)}$& $\varepsilon ~(\%)$ &
~~~~~~$\sigma_{\textnormal{\tiny MC}}$
$\pm$ $\varepsilon$ $\textnormal{(nb)}$ &$\varepsilon ~(\%)$ \\ && &&\\
\hline \hline &&&&\\ $gg \rightarrow 2g$ & (0.46572 $\pm$ 0.00258)$\times 10^{4}$&
~~0.5&  (0.46849 $\pm$ 0.00308)$\times 10^{4}$ &~~0.6\\
  $gg \rightarrow 3g$ & (0.15040 $\pm$ 0.00159)$\times 10^{3}$& ~~1.0
&(0.15127 $\pm$ 0.00110)$\times 10^{3}$& ~~0.7\\
 $gg \rightarrow 4g$  & (0.11873 $\pm$ 0.00224)$\times 10^{2}$
& ~~1.9& (0.12116 $\pm$ 0.00134)$\times 10^{2}$& ~~1.1\\
$gg \rightarrow 5g$ & (0.10082 $\pm$ 0.00198)$\times 10^{1}$
& ~~1.9&  (0.09719 $\pm$ 0.00142)$\times 10^{1}$& ~~1.5\\
 $gg \rightarrow 6g$ & (0.74717 $\pm$ 0.01490)$\times 10^{-1}$
& ~~2.0& (0.76652 $\pm$ 0.01862)$\times 10^{-1}$& ~~2.4  \\&&&&\\
\hline &&&&\\$gg \rightarrow  u\bar{u}$ &
(0.36435 $\pm$ 0.00199)$\times 10^{2}$
& ~~0.5 & (0.36619 $\pm$ 0.00132)$\times 10^{2}$  & ~~0.4\\
 $gg \rightarrow g u\bar{u}$ &  (0.35768 $\pm$ 0.00459)$\times 10^{1}$
& ~~1.3& (0.35466 $\pm$ 0.00291)$\times 10^{1}$ & ~~0.8\\
  $gg \rightarrow 2g u\bar{u}$ &  (0.49721 $\pm$ 0.00758)$\times 10^{0}$
& ~~1.5& (0.50053 $\pm$ 0.00725)$\times 10^{0}$& ~~1.4\\
  $gg \rightarrow 3g u\bar{u}$
& (0.50598 $\pm$ 0.01441)$\times 10^{-1}$& ~~2.8&
  (0.52908 $\pm$ 0.01264)$\times 10^{-1}$ & ~~2.4\\
  $gg \rightarrow 4g u\bar{u}$
& (0.51549 $\pm$ 0.02017)$\times 10^{-2}$& ~~3.9&
 (0.51581 $\pm$ 0.01245)$\times 10^{-2}$   &~~2.4\\
&&&&\\ \hline &&&&\\$gg \rightarrow   c\bar{c}c\bar{c}$ &
(0.25190 $\pm$ 0.00528)$\times 10^{-2}$ &  ~~2.1 &
(0.24903 $\pm$ 0.00373)$\times 10^{-2}$  & ~~1.5\\
  $gg \rightarrow  g c\bar{c}c\bar{c}$ &
(0.60196 $\pm$ 0.01908)$\times 10^{-3}$   & ~~3.2&
(0.58817 $\pm$ 0.00926)$\times 10^{-3}$ & ~~1.6\\
  $gg \rightarrow 2g c\bar{c} c\bar{c} $ &
(0.95682 $\pm$ 0.03441)$\times 10^{-4}$ & ~~3.6&
(0.92212 $\pm$ 0.02485)$\times 10^{-4}$ &~~2.7\\
&&&&\\ \hline\hline
\end{tabular}
\end{center}
\end{table}
\begin{table}[h!]
\newcommand{\lstrut}{{$\strut\atop\strut$}}
\caption {\it Comparison of the computational time for the squared
matrix element calculation for processes with gluons and/or quarks
with up to two $q\bar{q}$ pairs,
$\textnormal{t}_\textnormal{\tiny EXACT}$ denotes the
time for the processes calculated with summation over all possible
colour flows (for colour ordered amplitudes),
while $\textnormal{t}_\textnormal{\tiny MC}$ is obtained
with  Monte Carlo summation over colours (for full amplitudes).
In the last column their ratio
is presented. Time values are given in some arbitrary units. }
\vspace{0.2cm} \label{timing}
\begin{center}
\begin{tabular}{l|l|l|l}\hline \hline
&&&\\
~~$\textnormal{Process}$ &
~~$\textnormal{t}^{\textnormal{\tiny CF}}_{\textnormal{\tiny EXACT}}$ &
~~~~$\textnormal{t}_\textnormal{\tiny MC}$&
$\textnormal{t}_\textnormal{\tiny EXACT}/\textnormal{t}_\textnormal{\tiny MC}$
  \\  &&&\\
\hline \hline &&&\\
$gg \rightarrow 2g$ & 0.315$\times 10^{0}$ & 0.554$\times 10^{0}$ &~~~~0.57\\
$gg \rightarrow 3g$ & 0.329$\times 10^{1}$ & 0.143$\times 10^{1}$ &~~~~2.30\\
$gg \rightarrow 4g$ & 0.383$\times 10^{2}$ & 0.372$\times 10^{1}$ &~~~10.29\\
$gg \rightarrow 5g$ & 0.517$\times 10^{3}$ & 0.105$\times 10^{2}$ &~~~49.24\\
$gg \rightarrow 6g$ & 0.987$\times 10^{4}$ & 0.362$\times 10^{2}$ & ~~272.65\\
&&&\\
\hline
&&&\\
$gg \rightarrow u\bar{u}$ &
0.260$\times 10^{0}$ & 0.466$\times 10^{0}$ & ~~~~0.56\\
$gg \rightarrow g u \bar{u}$  &
0.196$\times 10^{1}$ & 0.123$\times 10^{0}$ & ~~~~1.59\\
$gg \rightarrow 2g u \bar{u}$ &
0.166$\times 10^{2}$ & 0.348$\times 10^{1}$ & ~~~~4.77\\
$gg \rightarrow 3g u \bar{u}$ &
0.171$\times 10^{3}$ & 0.129$\times 10^{2}$& ~~~13.25\\
$gg \rightarrow 4g u \bar{u}$ &
0.197$\times 10^{4}$& 0.307$\times 10^{2}$  & ~~~64.17\\
&&&\\
\hline
&&&\\
$gg \rightarrow c \bar{c}c \bar{c}$
& 0.697$\times 10^{1}$ & 0.605$\times 10^{1}$ & ~~~~1.15\\
$gg \rightarrow g c \bar{c}c \bar{c}$  &
0.568$\times 10^{2}$ & 0.217$\times 10^{2}$& ~~~~2.62\\
$gg \rightarrow 2g c \bar{c}c \bar{c}$ &
0.619$\times 10^{3}$& 0.401$\times 10^{2}$& ~~~15.44\\
&&&\\ \hline\hline
\end{tabular}
\end{center}
\end{table}

\begin{table}[h!]
\newcommand{\lstrut}{{$\strut\atop\strut$}}
\caption {\em  Results for the total cross section  for processes
with gluons, $Z$, $W^{\pm}$ and quarks with up to  $q\bar{q}$ pairs
for the higher number of external partons.
$\sigma_{\textnormal{{\tiny MC}}}$ corresponds to
 Monte Carlo summation over colours. }
\vspace{0.2cm}
\label{tab2}
\begin{center}
\begin{tabular}{l|l|l}
\hline \hline &&\\
~~~~~~$\textnormal{Process}$ &
~~~~~~$\sigma_{\textnormal{\tiny MC}}$
$\pm$ $\varepsilon$ $\textnormal{(nb)}$ &$\varepsilon ~(\%)$ \\  &&\\
\hline \hline &&\\
$gg \rightarrow 7g $
&(0.53185 $\pm$ 0.01149)$\times 10^{-2}$ &~~2.1\\
$gg \rightarrow 8g $
& (0.33330 $\pm$ 0.00804)$\times 10^{-3}$ & ~~2.4 \\
$gg \rightarrow 9g $
& (0.13875 $\pm$ 0.00430)$\times 10^{-4}$ & ~~3.1 \\
&&\\
\hline &&\\
 $gg \rightarrow 5g u\bar{u}$
&(0.38044 $\pm$ 0.01096)$\times 10^{-3}$ &~~2.8\\
$gg \rightarrow 3g c\bar{c} c\bar{c} $
& (0.95109 $\pm$  0.02456)$\times 10^{-5}$   & ~~2.6 \\
  $gg \rightarrow 4g c\bar{c} c\bar{c} $
& (0.81400 $\pm$ 0.02583)$\times 10^{-6}$   & ~~3.2 \\
&&\\
\hline &&\\
$gg \rightarrow Z u \bar{u} gg$
& (0.18948 $\pm$ 0.00344)$\times 10^{-3}$   & ~~1.8 \\
$gg \rightarrow  W^+ \bar{u} d gg $
& (0.62704 $\pm$ 0.01458)$\times 10^{-3}$   & ~~2.3 \\
$gg \rightarrow  ZZ u \bar{u} gg$
& (0.16217 $\pm$ 0.00420)$\times 10^{-6}$   & ~~2.6 \\
$gg \rightarrow W^+ W^- u \bar{u} gg$
& (0.27526 $\pm$ 0.00752)$\times 10^{-5}$   & ~~2.7 \\
&&\\
\hline &&\\
$d\bar{d} \rightarrow Z u \bar{u} gg$
& (0.38811 $\pm$ 0.00569)$\times 10^{-5}$   & ~~1.5 \\
$d\bar{d} \rightarrow  W^+ \bar{c} s gg $
& (0.18765 $\pm$ 0.00453)$\times 10^{-5}$   & ~~2.4 \\
$d\bar{d} \rightarrow  ZZ gg gg$
& (0.99763 $\pm$ 0.02976)$\times 10^{-7}$   & ~~2.9 \\
$d\bar{d} \rightarrow W^+ W^- gg gg$
& (0.52355 $\pm$ 0.01509)$\times 10^{-6}$   & ~~2.9\\
&&\\\hline\hline
\end{tabular}
\end{center}
\end{table}

In this section numerical results for multi-parton production at the
LHC are presented. As we will show the Monte Carlo summation  over
colours, speeds up the calculation substantially, compared to the one
based on explicit summation, especially for processes with a large
number of colors, {\it i.e.} $gg\to ng$ with $n > 6$.

The centre of mass energy was chosen to be $\sqrt{s}=14$
$\textnormal{TeV}$.  In order to remain far from collinear and
soft singularities
and to simulate as much as possible the experimentally relevant
phase-space regions, we have chosen the following cuts:
\begin{equation}
\label{cuts}
p_{T_{i}} > 60 ~\textnormal{GeV}, ~~~~ |y_{i}|<2.5, ~~~~ \Delta R_{ij} > 1.0
\end{equation}
for each pair of outgoing partons $i$ and $j$. Here $p_{T_{i}}$
and $y_{i}$ are  the transverse momentum and rapidity of a parton respectively
defined as:
\begin{equation}
p_{T_{i}}=\sqrt{p_{x_{i}}^{2}+p_{y_{i}}^{2}}, ~~~~~y_{i}=\frac{1}{2}\ln
\left( \frac{E_{i}+p_{z_{i}}}{E_{i}-p_{z_{i}}} \right).
\end{equation}
In practice for massless quarks the rapidity is often replaced by
the pseudorapidity variable
 $\eta=-\ln \tan (\theta/2)$, where $\theta$ is the angle from the beam
direction  measured directly in the detector.
The last variable is  $\Delta R_{ij}$ which is
the radius of the cone of the parton defined as
\begin{equation}
\Delta R_{ij}=\sqrt{(\Phi_{i}-\Phi_j)^{2}+(\eta_{i}-\eta_j)^{2}}
\end{equation}
with azimuthal angle $\Delta \Phi_{ij}=\Phi_{i}-\Phi_j \in (0,\pi)$
\begin{equation}
\Delta \Phi_{ij}=\arccos \biggl( \frac{p_{x_{i}}p_{x_{j}}+p_{y_{i}}p_{y_{j}}}
{p_{T_{i}} p_{T_{j}} } \biggr).
\end{equation}
Quarks are treated as massless. All results are obtained with a
fixed strong coupling constant ($\alpha_s$=0.13).
For the parton structure functions, we
used  {\tt CTEQ6L1 PDF}'s parametrisation \cite{Pumplin:2002vw,Stump:2003yu}.
For the phase space generation we used  the
algorithm described in the Appendix,  whereas in several cases
results were cross checked with
\begin{itemize}
\item {\tt PHEGAS} \cite{Papadopoulos:2000tt}, which automatically constructs
 mappings of all possible peaking structures of a given
scattering process and uses self-adaptive procedures like
multi-channel optimisation \cite{Kleiss:1994qy}. {\tt PHEGAS}
exhibits a high efficiency especially in the case of $n(+\gamma)-$fermion
production in $e^{+}e^{-}$ collisions as long as the number of external
particles is smaller than  ${\cal O}(8)$.
\item {\tt HAAG}\cite{Hameren:2002tc},
which efficiently maps the so called ``antenna momentum structures''
typically occurring in the QCD amplitudes. {\tt HAAG}
becomes less efficient for
processes with higher  number of particles, more than ${\cal O}(8)$,
as is the case for any multi-channel  generator, where the computational
complexity problem arises when the number of channels increases.
\item {\tt RAMBO} \cite{Kleiss:1985gy}, a flat phase-space generator
{\tt RAMBO} generates the momenta distributed uniformly in phase
space so that a large number of events is needed to integrate the
integrands to acceptable precision, which results to a rather low
computational efficiency.
\end{itemize}

Total rates and various kinematical distributions are examined.
The algorithm described in the previous sections as well as in
Ref.\cite{Papadopoulos:2005vg,Papadopoulos:2005jv} has been used
to compute total cross sections for many
parton production processes. We give the
result with summation over all possible colour configurations, called 'exact',
$\sigma_{\textnormal{{\tiny EXACT}}}$,  as well as the result obtained
with  Monte Carlo summation over colours
$\sigma_{\textnormal{{\tiny MC}}}$.

In case of the explicit summation both colour flow and colour
configuration decomposition have been used to cross check results.
As far as helicity summation is concerned, a Monte Carlo over
helicities is applied.

The results presented for the total cross sections, have been
obtained for $10^6$ Monte Carlo points passing the selection cuts
given by Eg.(\ref{cuts}). In the Tab.~\ref{tab1}. the results for
the total cross section for processes with gluons and/or quarks with
up to two $q\bar{q}$ pairs are presented. All cross sections are in
agreement within the error. For the same number of accepted events
the results with the Monte Carlo summation over colours can be
obtained much faster and the error is at the same level compared to
the 'exact' results for the same processes.

In the Tab.~\ref{timing}. a comparison of the computational time for
squared matrix element calculations for processes with gluons and/or quarks
with up to two $q\bar{q}$ pairs is presented. In each case
$\textnormal{t}_\textnormal{\tiny EXACT}$ means time obtained for
the processes calculated
with the summation over all possible
colour flows (for colour ordered amplitudes),
while $\textnormal{t}_\textnormal{\tiny MC}$ is obtained
with  Monte Carlo summation over colours (for full amplitudes).

In the next Tab.~\ref{tab2}. the results
for the total cross section for processes with gluon, $Z$, $W^{\pm}$ and
quarks with up to  $q\bar{q}$ pairs for a larger number of external partons
are also presented.
\begin{figure}[h!]
\begin{center}
\epsfig{file=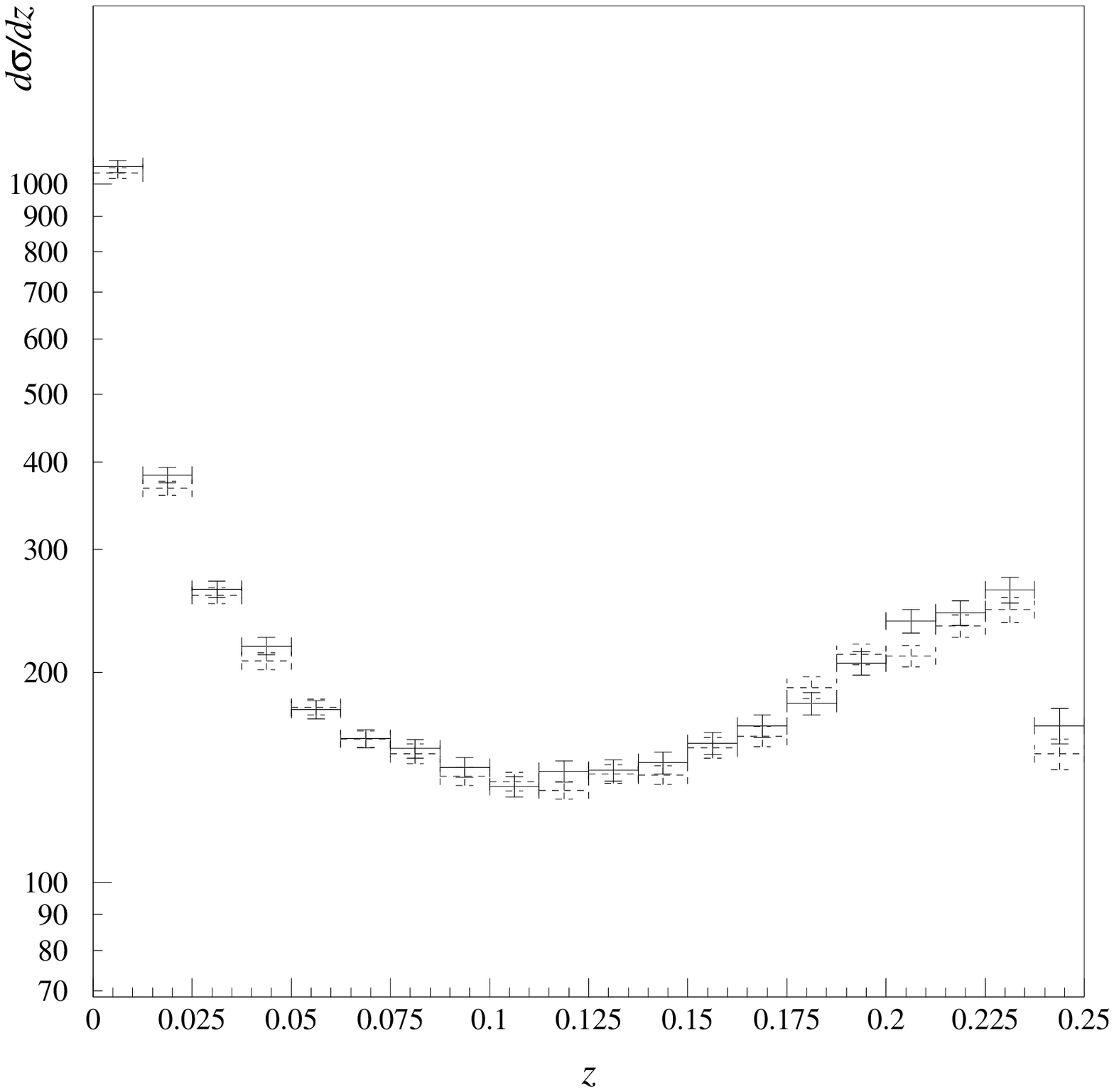,width=75mm,height=65mm}
\epsfig{file=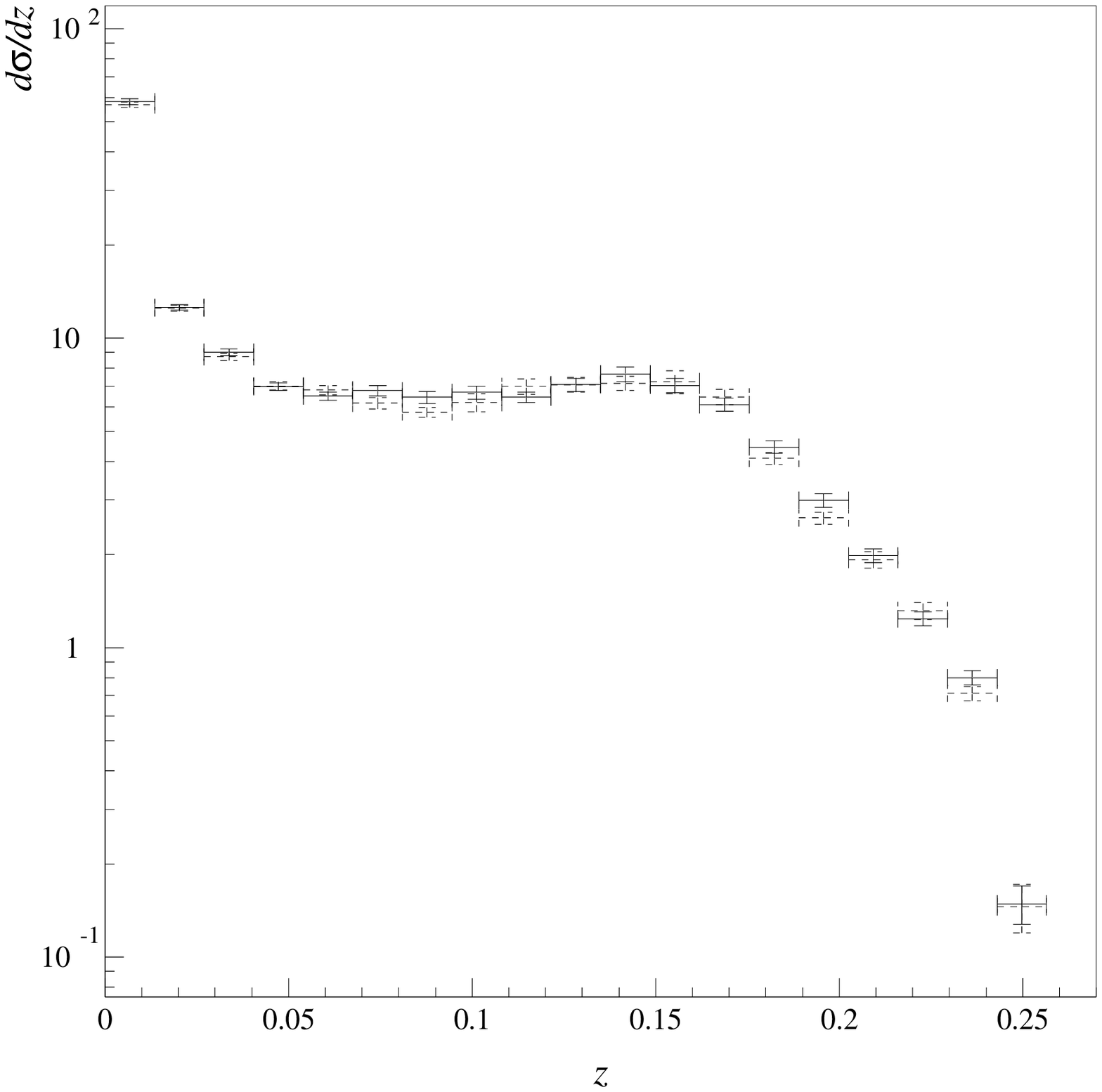,width=75mm,height=65mm}
\end{center}
\caption{\it Distribution in $z=|{\cal M}_{\textnormal{\tiny one}}
|^2/\sum_{i}^{all} |{\cal M}_{i}|^{2}$,
where $|{\cal M}_{\textnormal{\tiny one}}|^2$
is the square matrix element for one particular colour
configuration normalised to the sum of all possible. The
left-hand side plot corresponds to the
 $gg \rightarrow 2g$ process, the right-hand
side one to the $gg \rightarrow 3g$.
Solid line crosses denote summation over all colour configurations
whereas dashed, the Monte Carlo summation. }
\label{4g_5g_flows}
\end{figure}
\begin{figure}[h!]
\begin{center}
\epsfig{file=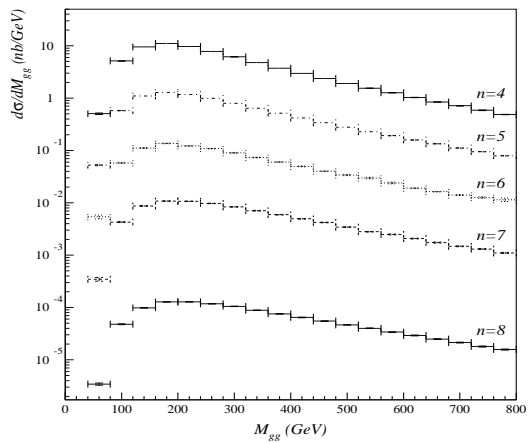,width=75mm,height=65mm}
\end{center}
\caption{\it Invariant mass distribution of 2 gluons in the multigluon
$2g \rightarrow ng$ process.}
\label{inv_mass}
\end{figure}
\begin{figure}[h!]
\begin{center}
\epsfig{file=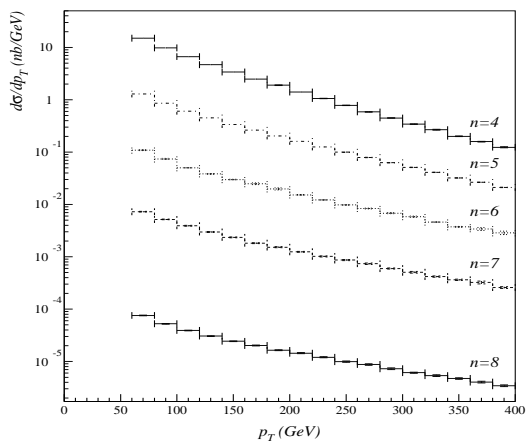,width=75mm,height=65mm}
\end{center}
\caption{\it  Transverse momentum distribution of a gluon in
the multigluon $2g \rightarrow ng$ process.}
\label{pt}
\end{figure}
\begin{figure}[h!]
\begin{center}
\epsfig{file=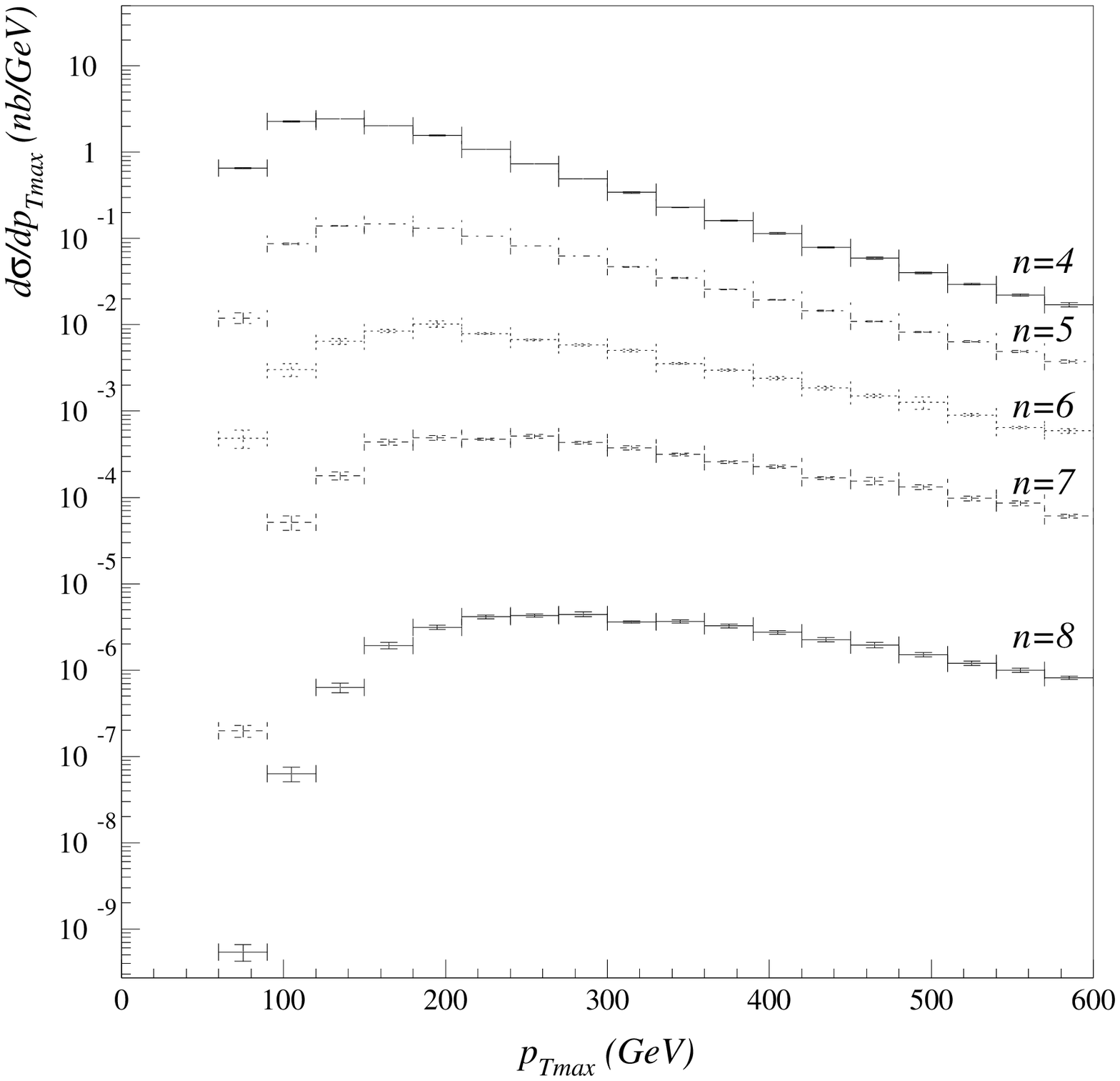,width=75mm,height=65mm}
\end{center}
\caption{\it Transverse momentum distribution of the hardest
 gluon  in the multigluon $2g \rightarrow ng$ process.}
\label{pt_max}
\end{figure}
\begin{figure}[h!]
\begin{center}
\epsfig{file=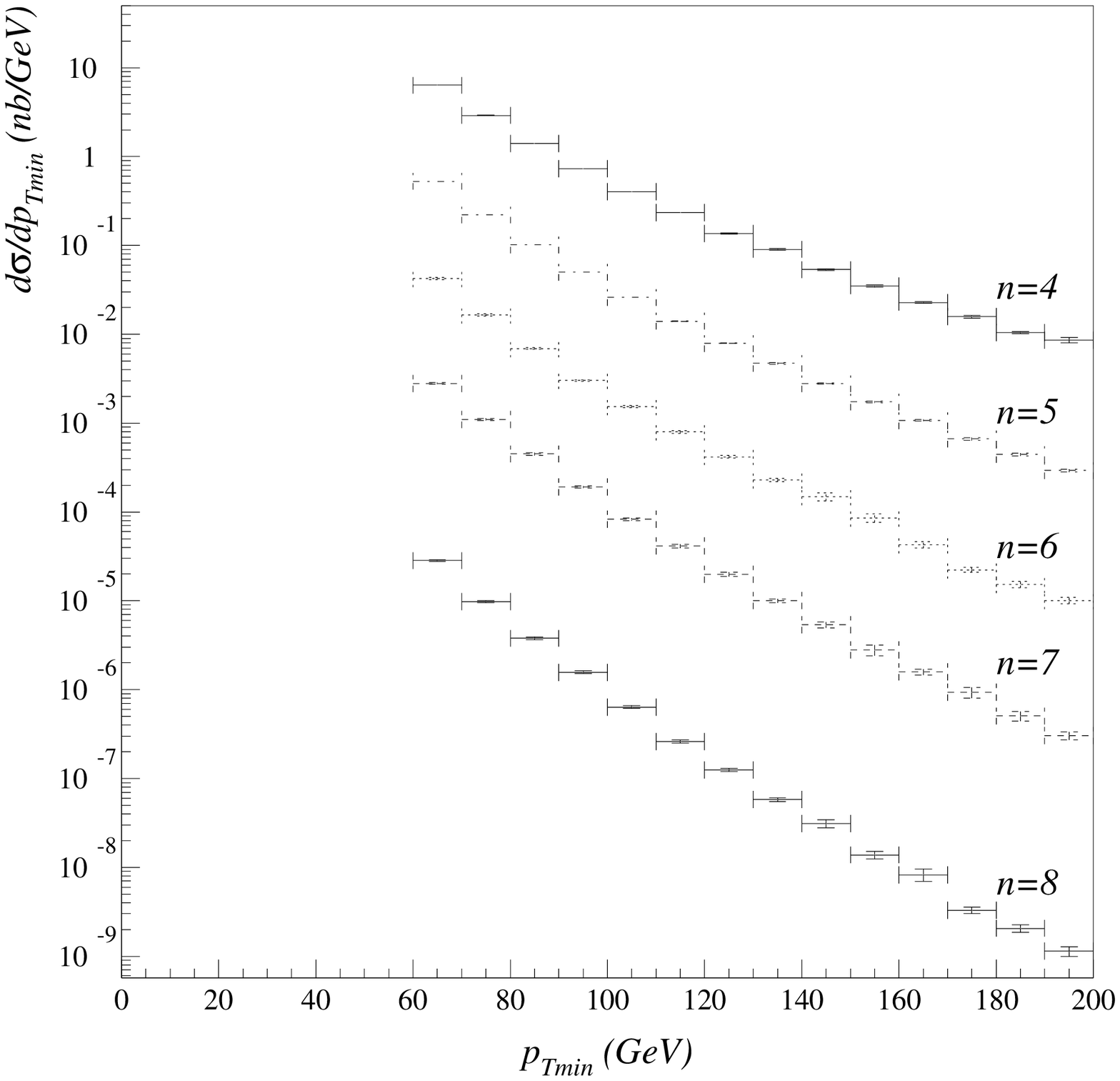,width=75mm,height=65mm}
\end{center}
\caption{\it Transverse momentum distribution of the softest  
gluon in the multigluon $2g \rightarrow ng$ process.}
\label{pt_min}
\end{figure}

In order to demonstrate that the Monte Carlo summation over colours
 does give the
same information on the colour connection structure of the process, we
examine  in Fig.~\ref{4g_5g_flows} the distribution of the following
variable,
\begin{equation}
z=\frac{|{\cal M}_{\textnormal{\tiny one}} |^2}{\sum_{i=1}^{all}
  |{\cal M}_{i}|^{2}}
\end{equation}
where $|{\cal M}_{\textnormal{\tiny one}}|^2$  is the square matrix
element for one particular colour connection or colour flow
configuration, normalised to the sum of all possible ones.  In case of
$g(1)g (2)\rightarrow g(3)g(4)$ process,  the following colour flow
$(1\rightarrow 3 \rightarrow 4 \rightarrow 2 \rightarrow 1)$ has been
plotted. This chain of
numbers shows how gluons are colour connected with each other.
For the $g(1) g(2) \rightarrow g(3) g(4) g(5)$
process in the Fig.~\ref{4g_5g_flows}, the colour flow
$(1 \rightarrow 5 \rightarrow 3 \rightarrow 2 \rightarrow 4 \rightarrow 1)$
is used.

The $z-$variable distributions can be used in order to extract
colour connection information needed by a parton shower calculation
on the event by event basis. The agreement between the 'exact' and
MC distributions implies  that we will get the same information on
the colour structure of the amplitude and that the merging of parton
level calculation with the parton shower evolution can be safely
performed in order to achieve a complete description of the fully
hadronised final states observed in   real experiments.

Finally, the distribution of the invariant mass  
of 2-partons, the transverse momentum distribution of 
 one parton as well as of the hardest and the softest parton for 
$gg\rightarrow ng$, $n=4,\ldots,8$  process is shown in
Fig.~\ref{inv_mass}.$-$Fig.~\ref{pt_min}.
Distributions of the  Monte Carlo over colours clearly demonstrate 
that this approach performs very well not only at the level of total rates
but also at the level of differential distributions.

\section*{Summary and Outlook}

In this work an efficient way for the computation of tree level
amplitudes for multi-parton processes in the Standard Model was
presented. The algorithm is based on the Dyson-Schwinger recursive
equations. We discussed how the summation over colour configurations
can be turned into a Monte Carlo summation, which proved to be more
efficient, especially for a large number of coloured partons.
Additionally, a set of typical results for total cross sections and
differential distributions has been given.  Moreover, a new algorithm
for phase-space generation has been presented and used.   The complete
package can be used to generate, efficiently and reliably, any process
with any number of external legs, for $n\le ~ 12$, in the Standard
Model.

Our future interest includes a systematic study of fully hadronic
final states in $p\bar{p}$ and $pp$ collisions  which requires the
merging of the parton level calculations with parton shower and
hadronization algorithms, {\it e.g} interfacing our package with
codes like  {\tt PYTHIA}\cite{Sjostrand:2000wi} or {\tt
HERWIG}\cite{Corcella:2000bw}.  The development of this kind of
multipurpose Monte Carlo generators  will certainly be of great
interest in the study of TeVatron, LHC and $e^{+}e^{-}$  Linear
Collider data.

\section*{Acknowledgments}

Work supported by the Polish State Committee for Scientific Research
Grants number 1 P03B 009 27 for years 2004-2005 (M.W.).  In
addition, M.W. acknowledges the Maria Curie Fellowship granted by
the European Community in the framework of the Human Potential
Programme under contract HPMD-CT-2001-00105 ({\it ``Multi-particle
production and higher order correction''}). The Greece-Poland
bilateral agreement {\it ``Advanced computer techniques for
theoretical calculations and development of simulation programs for
high energy physics experiments''}  ~is also acknowledged.

\section*{Appendix}

In case of scattering of two hadrons it is useful to describe the
final state in terms of  transverse momentum $p_{T}$, azimuthal angle
$\phi$ and rapidity $y$ variables. These variables transform simply
under longitudinal boosts which is useful  in case of
the parton-parton scattering  where the centre of mass
system is boosted with respect to that of the two incoming hadrons.
In terms of  $p_{T}$, $\phi$ and $y$ the four-momentum of a massless
particle can be written as
\begin{equation}
p^{\mu}=(E,p^{x},p^{y},p^{z})=(p_{T}\cosh y,p_{T}\cos \phi, p_{T}\sin
\phi,p_{T}\sinh y)
\end{equation}
It is much more natural to express the phase
space volume for a system of $n$ particles
\begin{equation}
V_{n}=\int \delta^{4} (
P-\sum_{i=1}^{n}p_{i})\prod_{i=1}^{n}d^{4}p_{i}
\delta(p_{i}^{2}-m^{2}_{i})\Theta(p^{0}_{i})
\end{equation}
using $p_{T}$, $\phi$ and $y$ variables
\begin{equation}
V_{n}=\int \delta^{4} ( P-\sum_{i=1}^{n}p_{i})\prod_{i=1}^{n}
p_{T_{i}} dp_{T_{i}} dy_{i} d\phi_{i}.
\end{equation}
To derive the above expression, which lead us to the Monte
Carlo algorithm, we follow the method presented in
Ref.~\cite{Kleiss:1985gy}. We start by defining the phase-space-like object
\begin{equation}
V_{0}=\int_{0}^{\infty}\left(\prod_{i=2}^{n}dk_{T_{i}}~P(k_{T_{i}})\right)
\int_{0}^{2\pi}\left(\prod_{i=2}^{n}d\phi_{i}\right)
\int_{-\infty}^{+\infty}
\left(\prod_{i=2}^{n}d{\bar y}_{i}~\Pi(\bar{y}_{i}) \right)
\label{V0}
\end{equation}
describing  a system of $n$ four momenta that are not constrained by
momentum conservation but occur with some weight functions
$P(k_{T_{i}})$, $\Pi({\bar y}_{i})$  which keeps the total volume
finite. In the next step we have to relate the new variables to the
physical ones  $p_{T_{i}}$, $y_{i}$ and $\phi_{i}$:
\[
V_{0}=\int_{0}^{\infty}\left(\prod_{i=2}^{n}dk_{T_{i}}~P(k_{T_{i}})\right)
\int_{0}^{2\pi}\left(\prod_{i=2}^{n}d\phi_{i}\right)
\int_{-\infty}^{+\infty}\left(\prod_{i=2}^{n}d{\bar y}_{i}~
\Pi(\bar{y}_{i})\right)
\]
\[
\int_{0}^{\infty}\left(\prod_{i=1}^{n}dp_{T_{i}}
~\delta(p_{T_{i}}-x k_{T_{i}})\right)
\int_{-\infty}^{+\infty}\left(\prod_{i=2}^{n}dy_{i}~
\delta(\bar{y}_{i}+y_{i-1}-y_{i})\right)
\]
\begin{equation}
\!\!\!\!\!\!\int_{0}^{\infty}dk_{T_{1}}\int_{0}^{2\pi}d\phi_{1}~
\delta(x\sum_{i=2}^{n} k_{T_{i}}\cos{\phi_{i}})~
\delta(x\sum_{i=2}^{n} k_{T_{i}}\sin{\phi_{i}})~{\cal J}_{1}
\end{equation}
\[
~~~~~~~~\int_{0}^{\infty}dx \int_{-\infty}^{+\infty}dy_{1}~
\delta(x\sum_{i=2}^{n}k_{T_{i}}\cosh y_{i}-E)~
\delta(x\sum_{i=2}^{n}k_{T_{i}}\sinh y_{i}-L)~{\cal J}_{2}.
\]
with  the Jacobians ${\cal J}_{1}$ and ${\cal J}_{2}$:
\begin{eqnarray}
&&{\cal J}_{1}=\left|
\frac{\partial(x\sum k_{T_{i}}\cos{\phi_{i}},x\sum
 k_{T_{i}}\sin{\phi_{i}})}{\partial(k_{T_{1}},\phi_{1})}\right|
=x^{2}k_{T_{1}}\\
&&{\cal J}_{2}=\left|
\frac{\partial(x\sum k_{T_{i}}\cosh{y_{i}}-E,x\sum
 k_{T_{i}}\sinh{y_{i}}-L)}{\partial(x,y_{1})}\right|=\frac{E^{2}-L^{2}}{x}.
\end{eqnarray}
$E$ and $L$ represents the energy and longitudinal parts of the initial
two particles. We proceed with integration where the different
arguments of the various $\delta$ functions were manipulated in
order to perform the  integral
\begin{eqnarray}
&&\int_{0}^{\infty}\prod_{i=1}^{n} dk_{T_{i}}~\delta(p_{T_{i}}-xk_{T_{i}})\\
&&\int_{-\infty}^{+\infty}\prod_{i=2}^{n}d{\bar y}_{i} ~\delta({\bar y}_{i}
+y_{i-1}-y_{i}).
\end{eqnarray}
We are left with
\begin{equation}
V_{0}= \int \delta^{4} ( P-\sum_{i=1}^{n}p_{i})\left(\prod_{i=1}^{n}
p_{T_{i}} dp_{T_{i}} dy_{i} d\phi_{i}\right)
 \left(\prod_{i=2}^{n}P\left(\frac{p_{T_{i}}}{x}\right)
\frac{1}{p_{T_{i}}}\Pi(\bar{y}_{i})\right)(E^{2}-L^{2})\frac{2^{n}}{x^{n}}dx.
\end{equation}
We are free to choose the distribution functions so that the
total volume is kept finite. The criterion used is to minimize the
variance, by taking into account the anticipated form of the
multi-parton matrix elements, so we introduce
\begin{equation}
P(x)=\frac{1}{a}\exp\left(\frac{-x}{a}\right),
~~~~~~~\Pi({\bar y})=\frac{\tanh(2\eta+ \bar{y})
+\tanh(2\eta-\bar{y})}{8\eta}
\label{weight_function}
\end{equation}
where $a>0$ and perform the integration over $dx$
\begin{equation}
\int_{0}^{\infty}\left(\prod_{i=2}^{n} \frac{1}{a}\exp\left(
\frac{-p_{T_{i}}}{xa}\right)
\frac{1}{p_{T_{i}}}\right)\frac{1}{x^{n}}dx=\left(\prod_{i=2}^{n}
\frac{1}{p_{T_{i}}} \right) \left( \sum_{i=2}^{n}p_{T_{i}}
\right)^{-n+1}\Gamma(n-1).
\end{equation}
We finally arrive at the formula
\begin{equation}
V_{0}= \int \delta^{4} ( P-\sum_{i=1}^{n}p_{i})\left(\prod_{i=1}^{n}
p_{T_{i}} dp_{T_{i}} dy_{i} d\phi_{i}\right) ~\times
\end{equation}
\[
\left(\prod_{i=2}^{n}
\frac{1}{p_{T_{i}}} \right) \left( \sum_{i=2}^{n}p_{T_{i}}
\right)^{-n+1}\Gamma(n-1)
\left(\prod_{i=2}^{n}\Pi(\bar{y}_{i})\right)(E^{2}-L^{2})~2^{n}.
\]
On the other hand if  we applied  Eq.(\ref{weight_function}) to the formula
 Eq.(\ref{V0}) we can find
\begin{equation}
\int_{0}^{\infty}\left( \prod_{i=2}^{n}dk_{T_{i}}\frac{1}{a}
\exp\left(\frac{-k_{T_{i}}}{a}\right)\right)=1
\end{equation}
\begin{equation}
\int_{-\infty}^{+\infty}\left(
\prod_{i=2}^{n}d\bar{y}_{i}\Pi(\bar{y}_{i})\right)=(8\eta)^{n-1}
\end{equation}
\begin{equation}
\int_{0}^{2\pi}\left(\prod_{i=3}^{n}d\phi_{i}\right)
=(2\pi)^{n-1}
\end{equation}
and
\begin{equation}
V_{0}=(2\pi\cdot8\eta)^{n-1}.
\end{equation}
The weight  of the event is given by
\begin{equation}
W=\frac{(2\pi\cdot8\eta)^{n-1}} {S_{n}}
\label{MC_weight}
\end{equation}
where
\[
S_{n}=\int dp_{T_{i}} dy_{i} d\phi_{i}
\left(\prod_{i=2}^{n}
\frac{1}{p_{T_{i}}} \right) \left( \sum_{i=2}^{n}p_{T_{i}}
\right)^{-n+1}\Gamma(n-1)
\left(\prod_{i=2}^{n}\Pi(\bar{y}_{i})\right)(E^{2}-L^{2})~2^{n}.
\]
In the next step we  translate this description into a Monte Carlo
procedure and generate independently $n$ variables $k_{T_{i}}$,
$\phi_{i}$ and $\bar{y}_{i}=y_{i}-y_{i-1}$ and
assuming that $\phi_{1}=0$ as well as $y_{1}=0$.
Using the symbol
$\rho_{i}$ to denote a random number uniformly distributed in $(0,1)$ we
do this as follows:
\begin{eqnarray}
k_{T_{i}} &=& -a\log\rho_{i},~~~~~~~~~~~~~~~i=2,\ldots,n\\
\phi_{i} &=& 2\pi\rho_{i},~~~~~~~~~~~~~~~~~~~~i=2,\ldots,n
\end{eqnarray}
where $a$ is a free parameter.
For $\bar{y}_{i}$ variable we proceed in few steps starting with
\begin{equation}
 F_{i}=\exp(4\eta(2\rho_{i}-1))~~~~~~~~~~~i=2,\ldots,n
\end{equation}
where $\eta$ is a free parameter and
\begin{equation}
\cosh \bar{y}_{i}=\frac{(F_{i}+1)\sinh 2\eta}{{\cal Z}_{i}}, ~~~~~~~
\sinh \bar{y}_{i}=\frac{(F_{i}-1)\cosh 2\eta}{{\cal Z}_{i}}
\end{equation}
where
\begin{equation}
{\cal Z}_{i}=\sqrt{2F_{i}\cosh4\eta -(1+F_{i}^{2})}.
\end{equation}
To complete the description of the algorithm we have to find an
expression for $k_{T_{1}}$, $\phi_{1}$ and $y_{1}$ variables. We
start by defining the transversal part of the $2,\ldots,n$ system as
follows
\begin{equation}
X\equiv \sum_{i=2}^{n}k_{T_{i}}\cos\phi_{i}, ~~~~~~~
Y\equiv \sum_{i=2}^{n}k_{T_{i}}\sin\phi_{i}.
\end{equation}
>From these equations we have the following relations for $k_{T_{1}}$  and
$\phi_{1}$ to be able to describe the total $n$ particle system
\begin{equation}
\cos\phi_{1}=-\frac{X}{k_{T_{1}}},
~~~~~~~~~~~~\sin\phi_{1}=-\frac{Y}{k_{T_{1}}}
\end{equation}
where
\begin{equation}
k_{T_{1}}=\sqrt{X^{2}+Y^{2}}.
\end{equation}
The total energy and total longitudinal part of the system are
represented by
\begin{eqnarray}
&&{\cal E}=k_{T_{1}}+k_{T_{2}}\cosh(\bar{y}_{2})+k_{T_{3}}\cosh(\bar{y}_{3}
+\bar{y_{2}})+ \ldots\\
&&{\cal L}=~~~~~~~~k_{T_{2}}\sinh(\bar{y}_{2})+k_{T_{3}}\sinh(\bar{y}_{3}
+\bar{y_{2}})+ \ldots
\end{eqnarray}
so $y_{1}$ is given by
\begin{eqnarray}
\cosh y_{1} &=&\frac{{\cal E}E-{\cal L}L}
{\sqrt{{\cal E}^{2}-{\cal L}^{2}}
\sqrt{E^{2}-L^{2}}},\\
\sinh y_{1}&=&\frac{-{\cal L}E+{\cal E}L}
{\sqrt{{\cal E}^{2}-{\cal L}^{2}}
\sqrt{E^{2}-L^{2}}},\\
x&=&\sqrt{\frac{E^{2}-L^{2}}{{\cal E}^{2}-{\cal L}^{2}}}.
\end{eqnarray}
Finally  to get the final four momenta $p^{\mu}_{i}$
the following transformations are used:
\begin{eqnarray}
&&E_{i}=xk_{T_{i}}\cosh y_{i}\\
&&p^{x}_{i}=xk_{T_{i}}\cos\phi_{i}\\
&&p^{y}_{i}=xk_{T_{i}}\sin\phi_{i}\\
&&p^{z}_{i}=xk_{T_{i}}\sinh y_{i}.
\end{eqnarray}
This completes the description of the algorithm we have
to supplemented it with the prescription for the weight of a
generated event which is given by Eq.(\ref{MC_weight}).

\providecommand{\href}[2]{#2}

\end{document}